# Signatures of three-dimensional photoinduced superconductivity in $YBa_2Cu_3O_{6.48}$

M. Rosenberg[1,*], D. Nicoletti[1,#], M. Buzzi[1], A. Iudica[1,2], C. Putzke[1],
Y. Liu[3], B. Keimer[3], A. Cavalleri[1,4,&]

[1] *Max Planck Institute for the Structure and Dynamics of Matter, 22761 Hamburg, Germany*
[2] *Physics Department, Politecnico di Milano, 20133 Milan, Italy*
[3] *Max Planck Institute for Solid State Research, 70569 Stuttgart, Germany*
[4] *Department of Physics, Clarendon Laboratory, University of Oxford, Oxford OX1 3PU, United Kingdom*

[*] *maor.rosenberg@mpsd.mpg.de* [#] *daniele.nicoletti@mpsd.mpg.de* [&] *andrea.cavalleri@mpsd.mpg.de*

# Abstract

**Optical excitation of large-amplitude apical oxygen phonon oscillations has been shown to renormalize the electronic properties of $YBa_2Cu_3O_{6+x}$, inducing a superconducting-like optical response above equilibrium $T_C$. All of the evidence collected so far has been based on the changes of the THz frequency *c*-axis response. In these measurements, the capacitive interlayer coupling was seen to transform into a superconducting-like inductive response. This assignment was strengthened by recent measurements of ultrafast magnetic field expulsion. Here, we report an experimental determination of the transient in-plane optical properties, which has so far been elusive due to the high equilibrium reflectivity and the need to evaluate minute changes in the optical response. We report the appearance of a photoinduced in-plane optical gap $2\Delta \simeq 30\ \mathrm{cm}^{-1}$ and a divergent imaginary conductivity, both consistent with photoinduced superconductivity. A global fit to these data suggests that in- and out-of-plane electronic properties never completely equilibrate during the dynamics.**

## I. Introduction

When cooled below the superconducting transition temperature, high-$T_C$ cuprates exhibit a characteristic terahertz-frequency optical response, reflecting coherent interlayer transport and condensation of normal state quasiparticles [1,2,3,4]. A representative set of optical properties is displayed in Fig. 1(a,b), where we show the complex optical conductivity of underdoped $YBa_2Cu_3O_{6+x}$ along the $c$ (out-of-plane) axis, for $T \gtrsim T_C$ (red) and $T \ll T_C$ (blue).

In the normal state, the real part of the optical conductivity, $\sigma_1(\omega)$, exhibits a flat and featureless spectrum below $\omega \simeq 100$ cm$^{-1}$ [Fig. 1(a)]. At higher frequencies a collection of absorption peaks is observed, assigned to infrared-active phonons, along with a broader contribution around 400 cm$^{-1}$, attributed to the so-called "transverse Josephson plasmon" [5,6,7,8]. For $T < T_C$, a depletion in $\sigma_1(\omega)$ occurs and the spectral weight condenses into a zero-frequency pole, indicative of dissipation-less transport. This is reflected in the imaginary part of the optical conductivity, $\sigma_2(\omega)$, as a $1/\omega$ divergence at low frequency [Fig. 1(b)], which allows the determination of the superfluid density from finite frequency measurements, as $\lim_{\omega \to 0} \omega\sigma_2(\omega) = \frac{n_S e^2}{m}$ (here $n_S$ is the superfluid density, $e$ the electron charge, and $m$ the electron mass).

In a series of recent experiments, femtosecond mid-infrared pulses have been used to resonantly excite apical oxygen phonon modes in underdoped $YBa_2Cu_3O_{6+x}$ [9,10,11,12,13,14,15,16]. These studies revealed a light-induced transient optical response featuring signatures of a finite superfluid density along the out-of-plane direction, as seen most directly in the transient $c$-axis $\sigma_2(\omega)$ which acquired the same $1/\omega$ behavior observed below $T_C$ [9,10,14,15,16]. Remarkably, this photoinduced response was observed all the way up to the pseudogap temperature, $T^*$ [10,14,15].

Representative results are reported in Figs. 1(c,d), where we show the complex optical conductivity of $YBa_2Cu_3O_{6.48}$ measured at $T = 100$ K $\simeq 2 \cdot T_C$ before (red) and after (blue) mid-infrared photoexcitation (here, the region covered by the pump spectrum is shaded in blue) [16]. However, this analogy did not apply to the real part of optical conductivity. In the transient state, the insulating low-frequency $\sigma_1(\omega)$ showed an increase [Fig. 1(c)], as opposed to a decrease upon cooling [Fig. 1(a)]. This response could be reproduced by fitting the data with a two-fluid model, in which a superconducting term coexisted with an overdamped Drude absorption accounting for hot and incoherent normal carriers [15,16,17,18,19].

In a recent work [16], both the enhancement of the superfluid density, $\omega\sigma_2(\omega)$, and the dissipative response of quasiparticles, $\sigma_1(\omega)$, were investigated by systematically tuning the duration and energy of the mid-infrared pump pulses, while keeping their peak field fixed. It was shown that the lifetime of the superconducting-like response coincided with the duration of the excitation pulse, up to at least $\tau \simeq 4$ ps. This allowed for a determination of well-defined optical properties over the entire probed range (as $1/\tau \simeq$ 0.25 THz), removing potential ambiguities in their low-frequency limit. It was also found that the photoinduced superfluid density reached values compatible with the zero-temperature equilibrium value for pulses made longer than the phonon dephasing time, while the dissipative component continued to grow with increasing pulse duration. An optimal regime of pump pulse durations was identified, for which the superconducting response was maximum and that of hot quasiparticles was minimized.

Overall, this series of experiments yielded a comprehensive picture for the out-of-plane superconducting-like optical response of photoexcited $YBa_2Cu_3O_{6+x}$, addressing in detail its dependence on doping, temperature, drive duration and photon energy. These results were also complemented by the recent discovery of a transient Meissner-like magnetic field expulsion under the same excitation conditions [20], as well as by work with second-harmonic probes that associated optically-driven superconductivity in $YBa_2Cu_3O_{6.48}$ with a parametric excitation of Josephson plasmons, which are overdamped above $T_C$ but would be made coherent by the phonon drive [21,22,23].

Nevertheless, a number of issues remain debated [24,25]. In the experiments discussed above, the penetration depth of the mid-infrared pump ($\approx$ 1 μm) was always shorter than that of the THz probe (5 − 10 μm). This mismatch was taken into account by modeling the sample as a multilayered photoexcited stack on top of an unperturbed bulk in order to obtain the optical response functions corresponding to an effective semi-infinite and homogeneously excited medium. This approach implies that, as the pump penetrates the material, its intensity is reduced and induces progressively weaker changes in the refractive index of the sample. However, for a given exponential decay of the pump intensity in the material depth, different refractive index decay profiles are possible, depending on the scaling of the pump-probe response with excitation fluence. Models with linear, square root dependence, or assuming a saturation profile were analyzed [26,27], all returning qualitatively similar changes in transient optical properties, although different in amplitude. In Fig. 1(c,d) we show the uncertainty associated with the specific choice of model as line thickness in $\sigma_1(\omega)$ and $\sigma_2(\omega)$. It becomes evident that the superconducting-like character of the response is preserved.

Another open issue concerns the optical response measured along the Cu-O layers. If the interlayer optical conductivity is compatible with dissipation-less transport [16] and magneto-optical measurements return indications of a transient Meissner effect [20], one should also expect a superconducting-like response for probe light polarized along the planes. This was not reported so far for underdoped $YBa_2Cu_3O_{6+x}$. Transient in-plane optical properties have been difficult to measure because of the high equilibrium reflectivity, from which small changes in the optical response determine the transient conductivity. Recent experiments in $La_{2-x}Ba_xCuO_4$, which reported a Josephson-like transient interlayer response for near-infrared optical excitation, did not detect any superconducting-like features in the in-plane response [28]. Yet in this work excitation was achieved with near-infrared pulses, which likely resulted in a sufficient non-equilibrium quasiparticle density to mask any sign of transient gapping. Measurements of the in-plane dynamics after excitation with optimized, mid-infrared optical pulses have so far been missing.

Here, we discuss a first investigation of the transient optical response of apical-oxygen-driven $YBa_2Cu_3O_{6.48}$ at THz frequencies along the direction parallel to the Cu-O layers. We report complex optical conductivities upon photoexcitation *qualitatively* compatible with transient superconductivity. However, unlike in the case of out-of-plane probing, the in-plane properties are *quantitatively* different from those found when cooling below $T_C$ at equilibrium. We also propose a global fitting of these three dimensional superconducting-like responses. Whilst the reliability of our fit is not completely certain, it suggests that the in-plane and out-of-plane quasiparticle and superfluid dynamics never equilibrate.

## II. Experiment

The equilibrium optical response of a single crystal of $YBa_2Cu_3O_{6.48}$ was determined with a Fourier transform infrared spectrometer. The absolute reflectivity was measured for reference at $T = 100$ K in quasi-normal-incidence geometry along the $a$ crystallographic axes in the $\approx 15 - 800\ \mathrm{cm}^{-1}$ range with the gold evaporation technique. By using literature data [29,30,31] for the higher frequency range, we could perform Kramers–Kronig transformations and retrieve the complex optical conductivity (see Supplemental Material [32] for more details and Ref. [33] for source data).

Another crystal from the same batch with similar dimensions was then mounted to expose a surface determined by the out-of-plane ($\parallel c$) and one of the in-plane ($\parallel a$) crystallographic directions. The crystal was photoexcited using mid-infrared pump pulses generated using an optical parametric amplifier coupled to a difference frequency generation stage [32]. These pump pulses were polarized along the $c$ axis and made to propagate through dispersive, optically polished NaCl plates before impinging on the sample. This yielded excitation pulses with the same spectral content but duration varying from 300 fs up to 2.5 ps.

Broadband THz probe pulses ($\approx 15 - 80\ \text{cm}^{-1}$), whose polarization was set to be parallel to the $a$ axis, were focused on the sample and detected by electro-optical sampling after reflection in a 500-μm-thick ZnTe (110) crystal, yielding the photoinduced changes in the low-frequency complex reflection coefficient $\tilde{r}(\omega)$ as a function of pump-probe time delay [32].

In contrast to $c$-axis probe experiments, in all in-plane measurements reported here the penetration depth of the excitation pulses ($\approx 1$ μm) was typically larger than that of the THz probe ($\approx 0.2 - 0.7$ μm). Therefore, the probe pulses sampled a homogeneously excited volume and the transient optical properties could then be extracted directly, without the need to consider any pump-probe penetration depth mismatch in the data analysis. In this case, the complex refractive index of the photo-excited material, $\tilde{n}(\omega,\tau)$, was directly retrieved from the Fresnel relation, $\tilde{n}(\omega,\tau) = \frac{1-\tilde{r}(\omega,\tau)}{1+\tilde{r}(\omega,\tau)}$, and from this, the transient complex optical conductivity, $\tilde{\sigma}(\omega,\tau) = \frac{\omega}{4\pi i}[\tilde{n}(\omega,\tau)^2 - \varepsilon_\infty]$.

For consistency, all data were also analyzed with the same multilayer models used for the out-of-plane response, yielding very similar results [32].

## III. Results

As shown in Fig. 2(a,b), in the direction parallel to the Cu-O planes ($\parallel a$) the equilibrium normal state response of underdoped $YBa_2Cu_3O_{6+x}$ is that characteristic of a metal with Drude-like conductivity [29,30,31]: The real part, $\sigma_1(\omega)$, takes on large values, exceeding $4{,}000\ \Omega^{-1}\text{cm}^{-1}$, and grows monotonically at low frequencies, approaching the DC value, while the imaginary part, $\sigma_2(\omega)$, exhibits a peak at finite frequency in correspondence of the carrier scattering rate, and becomes vanishingly small toward zero frequency. In addition, a collection of narrow absorption peaks is observed for $\omega \gtrsim 100\ \text{cm}^{-1}$, which are assigned to infrared-active phonons, only partially screened by the presence of mobile carriers.

In the superconducting state below $T_C$ (blue curves) the low-frequency $\sigma_1(\omega)$ is strongly depleted for $\omega \lesssim 250\ \text{cm}^{-1}$, with a residual quasiparticle peak indicative of a $d$-wave superconducting gap [29,30,31]. Correspondingly, $\sigma_2(\omega)$ exhibits a $1/\omega$-like divergence at low frequencies, which is a signature of dissipation-less transport [1].

The same type of equilibrium response has also been reported along the $b$ crystallographic axis, albeit with higher values of the normal state conductivity [34]. However, the $b$-axis electrodynamics is dominated by the Cu-O chains, an element specific to cuprates of the $YBa_2Cu_3O_{6+x}$ family and absent in other compounds. Here, we focus our study exclusively on the response along the $a$ axis.

In Fig. 2(c,d) we report representative results of our pump-probe experiment, in which we photoexcited $YBa_2Cu_3O_{6.48}$ at $T = 100\ \text{K} \simeq 2 \cdot T_C$ with 300-fs long, $c$-polarized mid-infrared pulses tuned to be resonant with apical oxygen vibrations, and then probed the $a$-axis complex optical conductivity throughout its dynamical evolution. Note that unlike for $c$-axis probe measurements, the absence of a pump-probe penetration depth mismatch results in transient optical properties that are effectively uncertainty-free, and essentially independent of the model used to reconstruct them [see line thickness in Fig.2(c,d)].

For the representative time delay shown here (blue curves), chosen to be at the peak of the response, we observe that $\sigma_1(\omega)$ reduces markedly from its equilibrium value, and develops a gap at low frequencies. Correspondingly, a $1/\omega$-like divergence appears in $\sigma_2(\omega)$ for $\omega \lesssim 30\ \text{cm}^{-1}$. Both of these features are suggestive of a superconducting-like response. However, while the photoinduced $\sigma_2(\omega)$ along the $c$ axis (see Fig. 1) was also quantitatively very close to that measured at equilibrium below $T_C$, the transient response along the $a$ axis is significantly smaller than the one measured in equilibrium when cooling below $T_C$. First, the photoinduced $\sigma_1(\omega)$ gap opens only at $\omega < 100\ \text{cm}^{-1}$, *i.e.* for frequencies much lower than the equilibrium below-$T_C$ gap [29]. Second, $\sigma_2(\omega)$ also develops a divergence, but the value of $\lim_{\omega\to 0} \omega\sigma_2(\omega)$, which would be proportional to the superfluid density, is lower by at least a factor of 10 than the equilibrium value. Note also that we find no evidence of a residual Drude peak, which instead is observed in equilibrium.

In Fig. 3 we report representative spectra for different pump pulse durations. These data show no qualitative change, but only a reduction in the amplitude of the response compared with those in Fig. 2, presumably due to the lower peak electric field. All complex conductivity spectra show gapping in $\sigma_1(\omega)$ and the appearance of a low-frequency divergence in $\sigma_2(\omega)$ following photoexcitation, which combined return a superconducting-like in-plane response (see Supplemental Material [32] for extended data sets).

## IV. Fitting procedure

The dynamical evolution of dissipation and superconducting coherence is summarized by plotting the transient $\sigma_1(\omega)$ spectral weight loss $\int_{20\ \mathrm{cm}^{-1}}^{75\ \mathrm{cm}^{-1}} \sigma_1(\omega)\,\mathrm{d}\omega$ and the strength of the inductive coupling as $\lim_{\omega\to 0} \omega\sigma_2(\omega)$, extracted from the individual THz spectra measured at each time delay (see also analysis in Ref. [16]).

Figure 4 displays these two figures of merit for a representative data set, for both probe polarizations. Upon photoexcitation the $\sigma_1(\omega)$ spectral weight is enhanced along the $c$- and reduced along the $a$-axis, relaxing then on a timescale that clearly exceeds the pump pulse duration of 2.5 ps [Figs. 4(a,c)]. A different evolution is observed instead in $\lim_{\omega\to 0} \omega\sigma_2(\omega)$, which appears to be finite along both crystallographic axes only while the system is driven [blue shading in Figs. 4(b,d)]. A first implication of this result is that the transient conductivities from which the data in Fig. 4 were derived are well defined only down to $1/\tau \simeq 0.4$ THz, a cutoff that is below the low-frequency limit covered by our probe spectrum. Therefore, any ambiguity in the definition of the optical properties does not apply to our data with long pump pulses.

We also note that for the data set reported here, $\lim_{\omega\to 0} \omega\sigma_2(\omega)$ reaches, upon photoexcitation, values close to the $c$-axis equilibrium superfluid density measured at low temperature (dashed horizontal line in Fig. 4(b), see also Ref. [16]). This is not found instead for the response along the planes, for which the transient "superfluid density" does not exceed a few percent of the equilibrium value [Fig. 4(d)].

We now turn to a more detailed analysis of transient optical properties and propose a model to fit the experimental data. In Ref. [16], it was discussed how the photoinduced response of the uncondensed quasiparticles and the "superfluid" were totally decoupled and showed up separately in the real and imaginary part of the optical conductivity, respectively. The transient spectra were fitted with a Josephson plasma model in the presence of a residual Drude conductivity, with the formula $\tilde{\sigma}_a(\omega) = \frac{\omega_J^2}{4\pi}\frac{i}{\omega} + \tilde{\sigma}_D(\omega)$. Here, $\omega_J$ is the Josephson plasma frequency, while $\tilde{\sigma}_D(\omega)$ is an overdamped Drude term which reproduces the temperature-dependent flat offset found in $\sigma_1(\omega)$.

By comparing the photoinduced $\sigma_1(\omega)$ spectra with those measured at equilibrium at various temperatures and assuming rapid thermalization, we estimate the effective "heating" of uncondensed quasiparticles for various pump fluences and time delays. Maximum apparent temperature raises up to $150 - 200$ K were extracted, in conjunction with a superconducting-like response in $\sigma_2(\omega)$, showing two completely decoupled dynamics of coherent and incoherent carriers.

This type of analysis is more complicated for the in-plane response. Along the planes the low-frequency $\sigma_2(\omega)$ is also dominated by the superconducting-like coherent response (see equilibrium spectra in Fig. 2(b)). However, the reduction in $\sigma_1(\omega)$ found upon photoexcitation can be attributed to two effects. On the one hand, a gapping in the optical conductivity due to the presence of a superconducting condensate is expected. On the other, as can be seen by comparing the spectra at equilibrium measured at $T = 120$ K and $T = 200$ K in Fig. 2(a), an increase in the quasiparticle scattering rate with temperature also leads to a reduction in the low-frequency $\sigma_1(\omega)$ in our measurement range.

We fitted the in-plane complex conductivity of $YBa_2Cu_3O_{6.48}$ for various time delays and under different excitation conditions with a two-fluid model. The first fluid describes the response of a superconductor with the Zimmermann model [35], *i.e.*, an extension of the Mattis-Bardeen model for superconductors of arbitrary purity, having the optical gap, $2\Delta$, as the fit parameter. The second fluid consists instead of unpaired residual quasiparticles [36], whose response is described by a Drude model with increased scattering rate, $\Gamma$, with respect to the equilibrium value, in order to account for carrier heating. In addition, to preserve the total number of carriers we introduced a filling fraction parameter, $f$, so that the total fit formula is expressed as: $\tilde{\sigma}_a(\omega) = f\tilde{\sigma}_{Zim}(\omega, 2\Delta) + (1-f)\,\frac{\omega_P^2}{4\pi}\frac{1}{\Gamma - i\omega}$.

Examples of these fits are given for specific datasets in Fig. 3 (see also Supplemental Material [32]). The divergence found in $\sigma_2(\omega)$ at low frequencies is entirely due to the superconducting term, while the reduction in $\sigma_1(\omega)$ is partly to be attributed to quasiparticles heating, which appears as a rigid downward shift and broadening, and partly to the opening of the gap in $\tilde{\sigma}_{Zim}$. Typical extracted parameters are $2\Delta \simeq 30\ \text{cm}^{-1}$, $\Gamma \simeq 2 \cdot \Gamma_{equil}$, and $f$ values around 20%.

This fitting procedure can be used to extract an equivalent temperature for the quasiparticles throughout the photoinduced dynamics in a similar way as reported in Ref. [16] for the out-of-plane response. Therein, by comparing the Drude parameters derived from the fits to the transient spectra with those performed on the conductivities at equilibrium for different temperatures (see also Fig. 2), we associated a quasi-temperature with each out-of-equilibrium conductivity at different time delays and excitation conditions [32].

We follow here the same procedure, and plot an example of these "apparent" temperature dynamics in Fig. 5 for data sets acquired with 2.5 ps pump pulse duration and both probe polarizations, taken under the same excitation conditions. The peak value of this estimated quasi-temperature is higher along the $c$ axis than along the planes, while the relaxation dynamics appears slower in the direction parallel to the planes. These differences are presumably a result of the inadequate fitting procedure and of the comparison with the equilibrium spectra. It is not unlikely that other parameters, such as the carrier effective mass (incorporated in the plasma frequency, $\omega_P$), are incorrectly assumed here to remain fixed during the dynamics.

## V. Conclusions

We reported a first study of the terahertz-frequency optical response of apical-oxygen-driven underdoped $YBa_2Cu_3O_{6+x}$ along the Cu-O planes, under the same excitation conditions for which an out-of-plane superconducting-like THz response and a transient magnetic field expulsion were observed. We measured the complex optical conductivity along the $a$ axis, representative of the in-plane response, at $T = 100\text{ K} \simeq 2 \cdot T_C$ for various excitation fluences and driving pulse durations, as a function of pump-probe time delay. Our findings are qualitatively similar to those expected for a transient superconducting response, with the opening of a gap in the real part of the optical conductivity, $\sigma_1(\omega)$, and the appearance of a low-frequency divergence in the imaginary part, $\sigma_2(\omega)$. By analyzing the dynamics of both the dissipative and coherent parts of the photoinduced response we observed a very similar evolution of the same quantities measured along the $c$ axis, finding that the coherent response in $\sigma_2(\omega)$ persists only as long as the drive is on. Using a two-fluid model we were then able to consistently fit the measured spectra and extract physical parameters such as the equivalent quasi-temperature of uncondensed quasiparticles.

Importantly, while the data reported along the $c$ axis showed quantitative matching with the superconducting response measured at equilibrium below $T_C$, the in-plane response yields a transient optical gap and "superfluid density" lower than those observed at equilibrium.

These results, when combined with previous experiments, suggest that the driven pseudogap phase of underdoped $YBa_2Cu_3O_{6+x}$, exhibits a three dimensional superconducting-like optical response, an observation which complements recent reports of an out-of-equilibrium Meissner effect. In this light, the explanations of magnetic field expulsion by large paramagnetism in a high temperature cuprate with preexisting equilibrium short range superconducting fluctuations appear less likely [37,38].

## Data Availability Statement

The data that support the findings of this article are openly available [33].

## Acknowledgments

We acknowledge support from the Deutsche Forschungsgemeinschaft (DFG, German Research Foundation) via the Cluster of Excellence CUI: Advanced Imaging of Matter (EXC 2056, project ID 390715994).

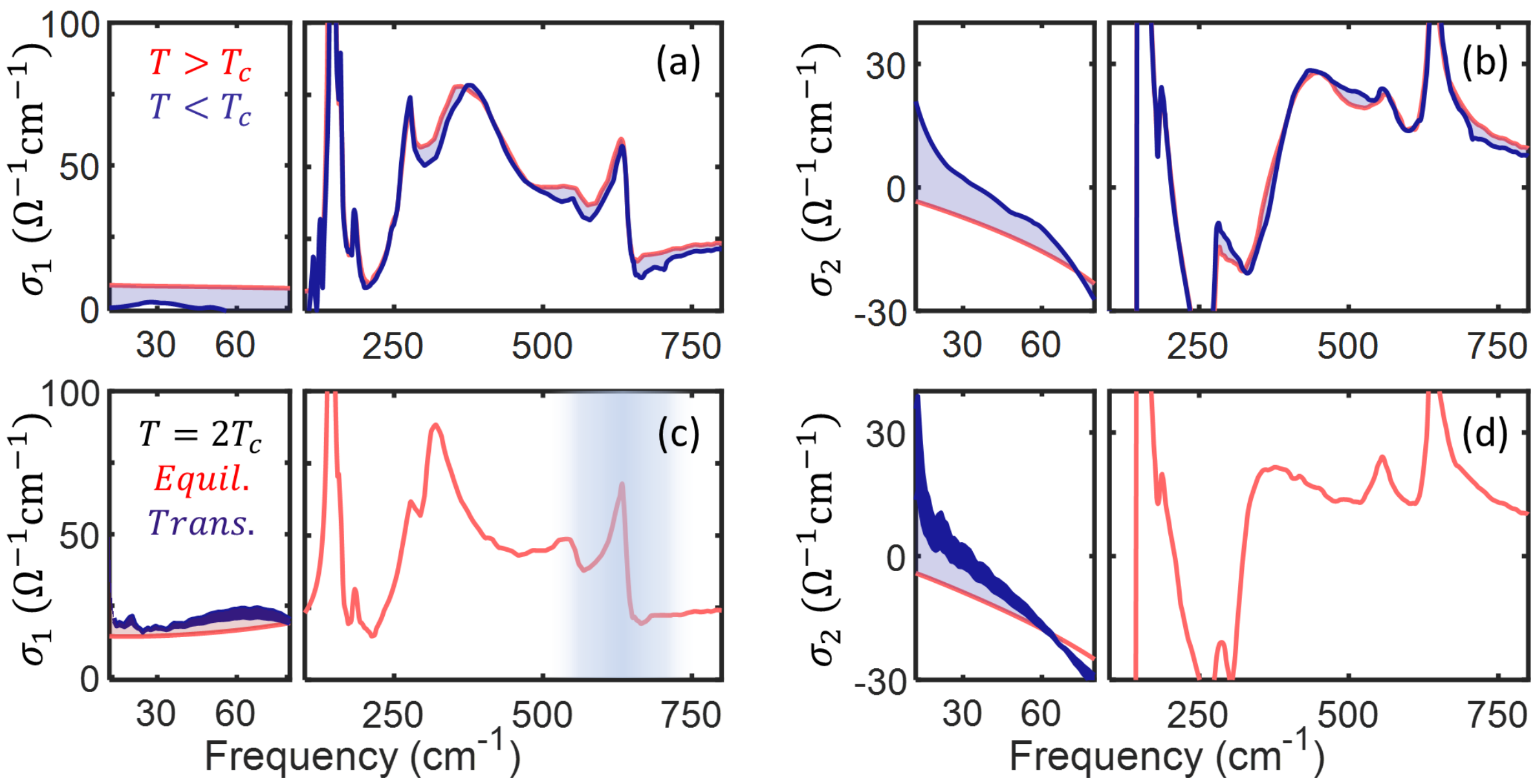

**Figure 1. (a,b)** Complex $c$-axis optical conductivity, $\sigma_1(\omega) + i\sigma_2(\omega)$, of underdoped $YBa_2Cu_3O_{6+x}$ across the equilibrium superconducting transition. Data at $T = 60\ K > T_C$ (red) and $T = 10\ K < T_C = 55\ K$ (blue) are reported [2,3,10,16]. Shaded blue regions represent the changes in the spectra across $T_C$. **(c,d)** Same quantities as in (a) and (b) measured in $YBa_2Cu_3O_{6.48}$ at $T = 100\ K \simeq 2 \cdot T_C$ before (red) and after (blue) photoexcitation with ~0.3 ps long mid-infrared pulses with a fluence of ~8 mJ/cm$^2$. Red and blue areas in (c) and (d) represent photoinduced changes in the spectra, while the blue shading around 600 cm$^{-1}$ in (c) refers to the frequency range covered by the mid-infrared pump. The thickness of the blue lines indicates the uncertainty associated with the specific choice of model to deal with the pump-probe penetration depth mismatch [32].

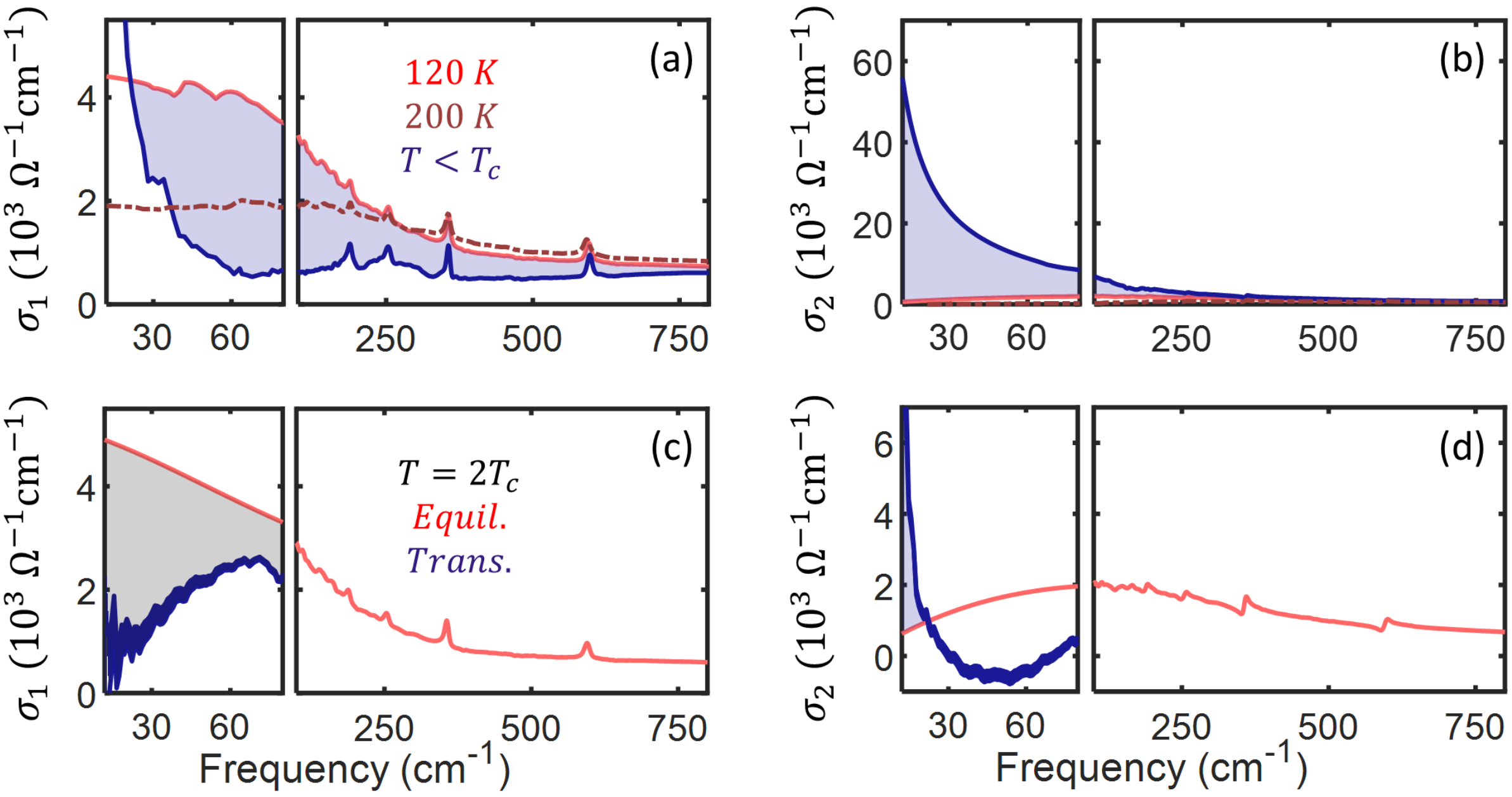

**Figure 2. (a,b)** Complex $a$-axis optical conductivity, $\sigma_1(\omega) + i\sigma_2(\omega)$, of underdoped $YBa_2Cu_3O_{6+x}$ across the equilibrium superconducting transition. Data at $T = 120$ K, 200 K $> T_C$ (red) and $T = 10$ K $< T_C = 55$ K (blue) are plotted [29]. Shaded blue regions represent the changes in the spectra across $T_C$. **(c,d)** Same quantities as in (a) and (b) measured in $YBa_2Cu_3O_{6.48}$ at $T = 100$ K $\simeq 2 \cdot T_C$ before (red) and after (blue) photoexcitation with ~0.3 ps long mid-infrared pulses with a fluence of ~8 mJ/cm$^2$. Shaded areas in (c) and (d) highlight the photoinduced gapping in $\sigma_1(\omega)$ and low-frequency divergence in $\sigma_2(\omega)$. The thickness of the blue lines indicates the uncertainty associated with the specific choice of model to extract the transient optical properties [32].

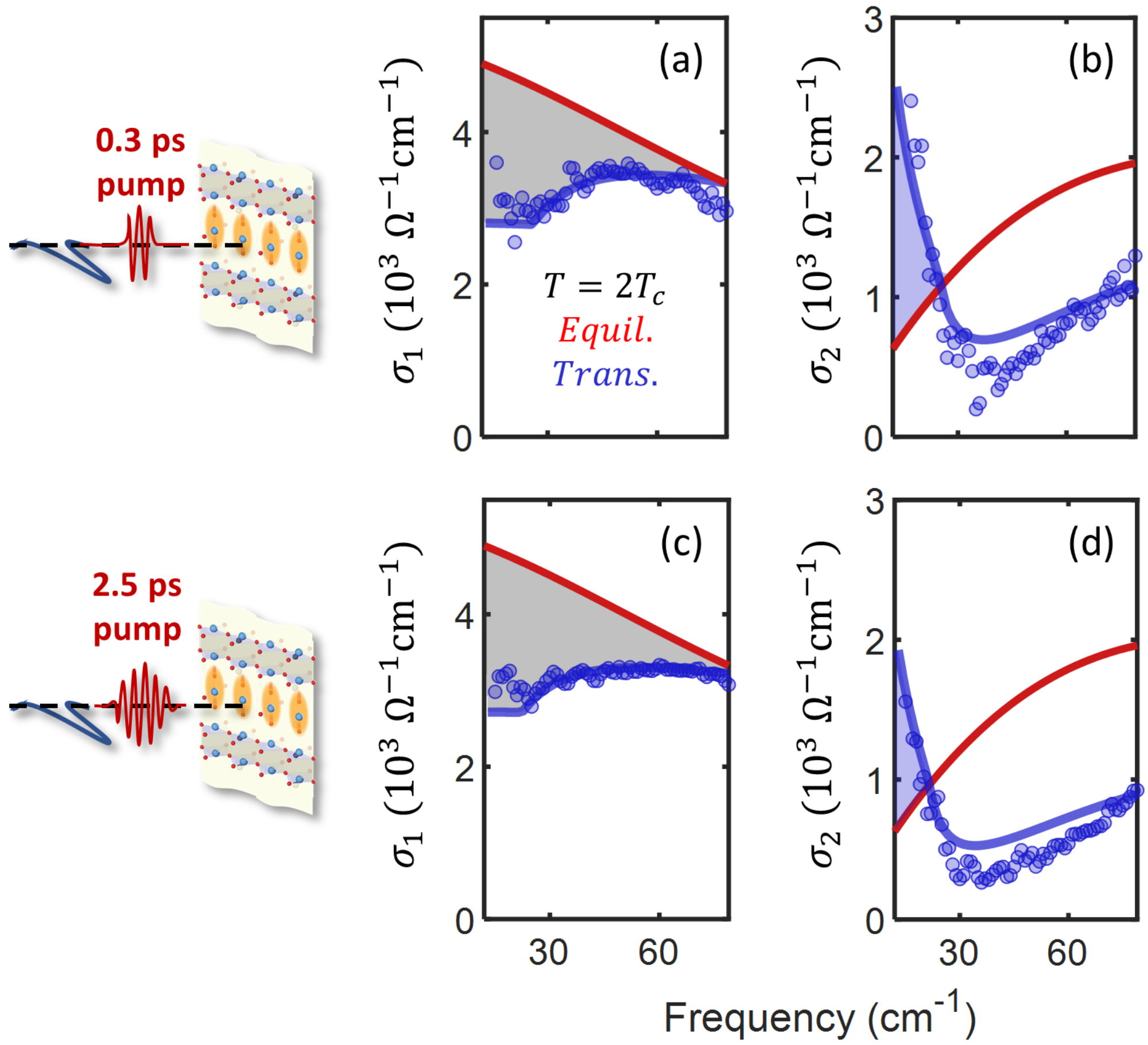

**Figure 3.** Complex $a$-axis optical conductivity, $\sigma_1(\omega) + i\sigma_2(\omega)$, measured in $YBa_2Cu_3O_{6.48}$ at $T =$ 100 K in equilibrium (red) and at one selected pump-probe time delay after photoexcitation (blue circles), corresponding to the peak of the coherent, superconducting-like response. We display data for different mid-infrared pulse durations (shown on the left), all taken with the same pump fluence of ~6.5 mJ/cm$^2$. Blue lines are fits to the spectra with the two fluid model described in the text [32], featuring a superconducting component described by the Zimmermann model [35] and a second fluid made of uncondensed quasiparticles, described by a simple Drude model [36].

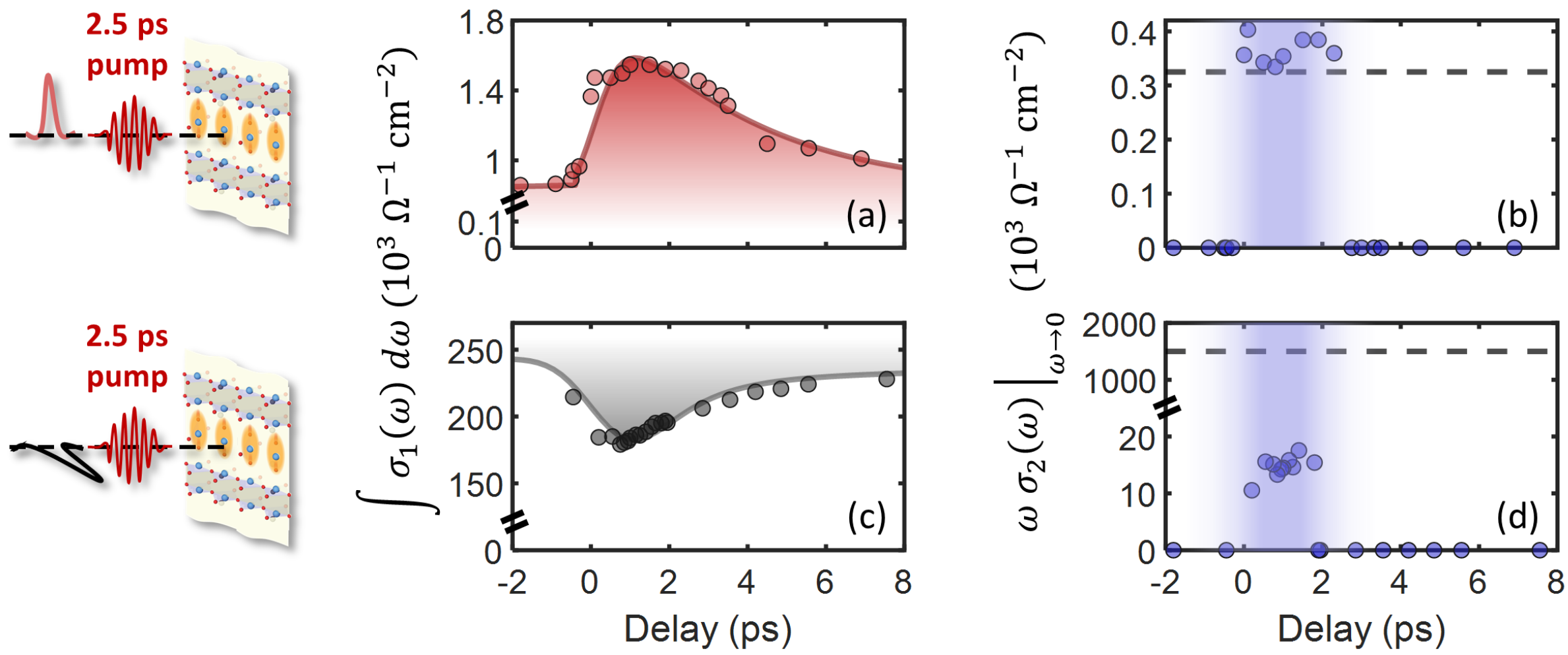

**Figure 4.** Dynamical evolution of the transient spectral weight, $\int_{20\ \mathrm{cm}^{-1}}^{75\ \mathrm{cm}^{-1}} \sigma_1(\omega)\,\mathrm{d}\omega$ and of the coherent superconducting-like response, $\lim_{\omega\to 0} \omega\sigma_2(\omega)$, as a function of pump-probe time delay, measured in $YBa_2Cu_3O_{6.48}$ at $T = 100$ K upon excitation with 2.5 ps long pulses, for both *c*- and *a*-axis probe polarizations (top and bottom panels, respectively). Data in (a,b) are replotted from Ref. [16], while those in (c,d) were acquired with a comparable pump fluence of $\sim$6.5 $\mathrm{mJ/cm^2}$. Full lines in (a,c) are fits with a finite rise time and an exponential decay, while the dashed horizontal lines in (b,d) indicate the 10 K equilibrium values of the superfluid density. The blue shaded areas in the same panels represent the time delay window for which a finite coherent response was detected.

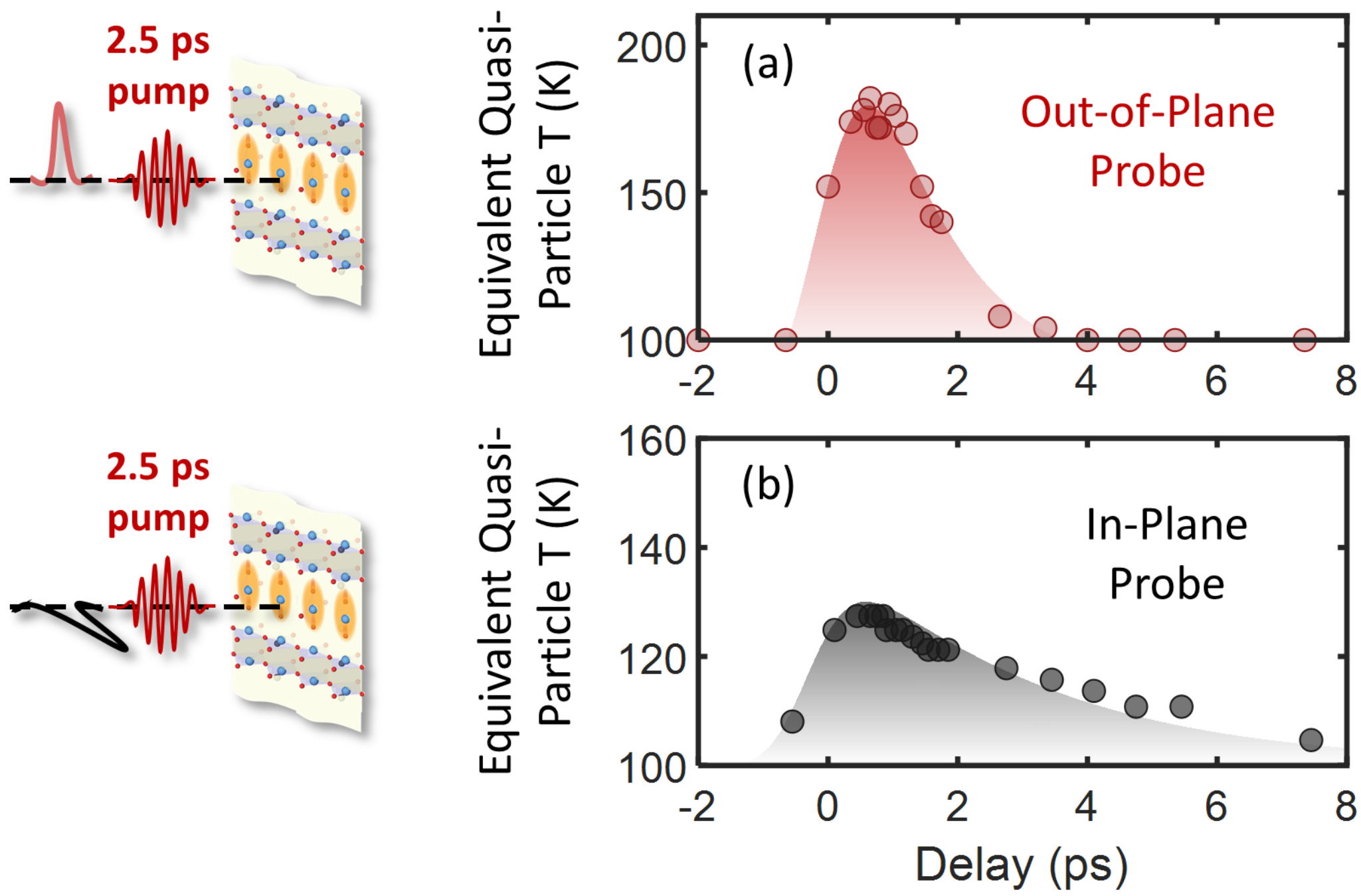

**Figure 5.** Equivalent quasi-temperatures for the uncondensed quasiparticles, extracted from fits to the complex optical conductivities measured in $YBa_2Cu_3O_{6.48}$ at a base temperature $T = 100$ K along both the $c$- and $a$-axis, as described in the text [32]. The data refer to measurements taken in the same setup with 2.5 ps drive pulses and a pump fluence of $\sim$6.5 mJ/cm$^2$.

# Signatures of three-dimensional photoinduced superconductivity in $YBa_2Cu_3O_{6.48}$

M. Rosenberg[1], D. Nicoletti[1], M. Buzzi[1], A. Iudica[1,2], C. Putzke[1], Y. Liu[3], B. Keimer[3], A. Cavalleri[1,4]

*[1] Max Planck Institute for the Structure and Dynamics of Matter, 22761 Hamburg, Germany*
*[2] Physics Department, Politecnico di Milano, 20133 Milan, Italy*
*[3] Max Planck Institute for Solid State Research, 70569 Stuttgart, Germany*
*[4] Department of Physics, Clarendon Laboratory, University of Oxford, Oxford OX1 3PU, United Kingdom*

# Supplemental Material

## S1. Sample growth and characterization

$YBa_2Cu_3O_{6.48}$ crystals, typically measuring 1.5 × 1.5 × 0.5 mm, were synthesized using Y-stabilized zirconia crucibles. The hole concentration in the Cu-O planes was tuned by regulating the oxygen content in the Cu-O chain layer through annealing in flowing oxygen followed by rapid quenching. A distinct superconducting transition at $T_C = 55$ K, with a transition width of approximately 2 K, was confirmed via dc magnetization measurements, as illustrated in Fig. S1. The correct orientation of the axes with respect to the exposed surfaces was verified with an X-ray diffractometer.

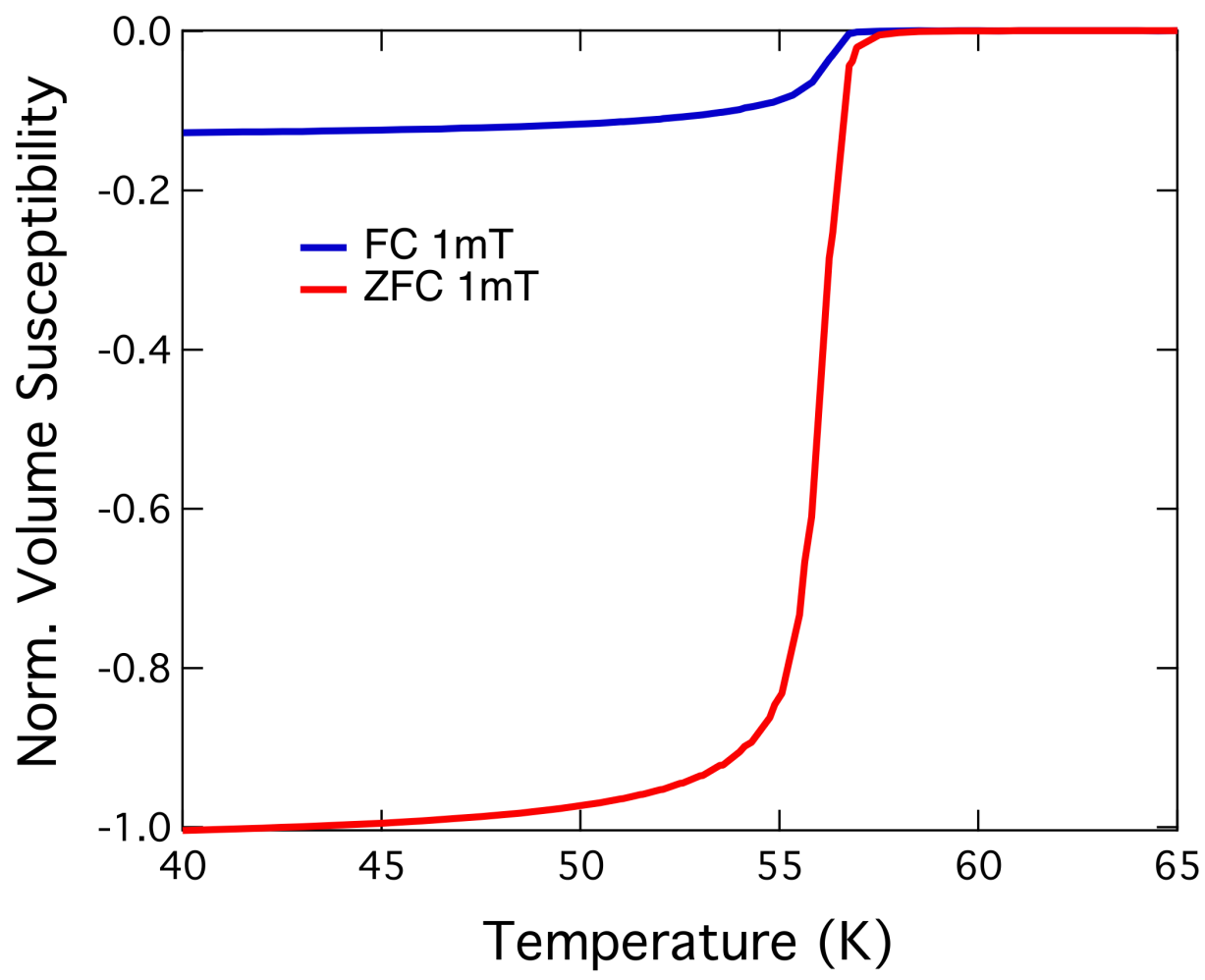

**Figure S1.** SQUID measurements of DC magnetization were performed to characterize the superconducting transition, using both zero-field-cooled (ZFC) and field-cooled (FC) protocols.

## S2. Equilibrium optical properties

The $YBa_2Cu_3O_{6.48}$ equilibrium optical properties were measured using a Bruker Vertex 80v interferometer. The sample was mounted on the tip of a cone-shaped holder, with the in-plane (*ab*) crystal surface exposed to the beam. The holder was installed on the cold finger of a He-flow cryostat, thus enabling to collect broadband infrared spectra in reflection geometry at different temperatures. A far-infrared polarizer was used to select the polarization parallel to the crystal $a$ axis. The spectra were then referenced against a thin gold film evaporated in-situ on the same sample surface.

The reflectivity curves obtained with this procedure, covering a range between $\sim 15-800\ \text{cm}^{-1}$, were extrapolated to $\omega \to 0$ with Drude fits (see also Section S5), and extended to higher frequencies with literature data[i,ii,iii]. This allowed us to perform Kramers-Kronig transformations and retrieve full sets of $a$-axis equilibrium response functions to be used as reference in our pump-probe experiment. Note that, in order to provide a smooth reference and avoid propagating noise from equilibrium optical properties, we used fits through the experimental reflectivity up to $\sim 100\ \text{cm}^{-1}$.

The reflectivity and complex optical conductivity spectra obtained with this procedure at $T = 100$ K are reported in Fig. S2.

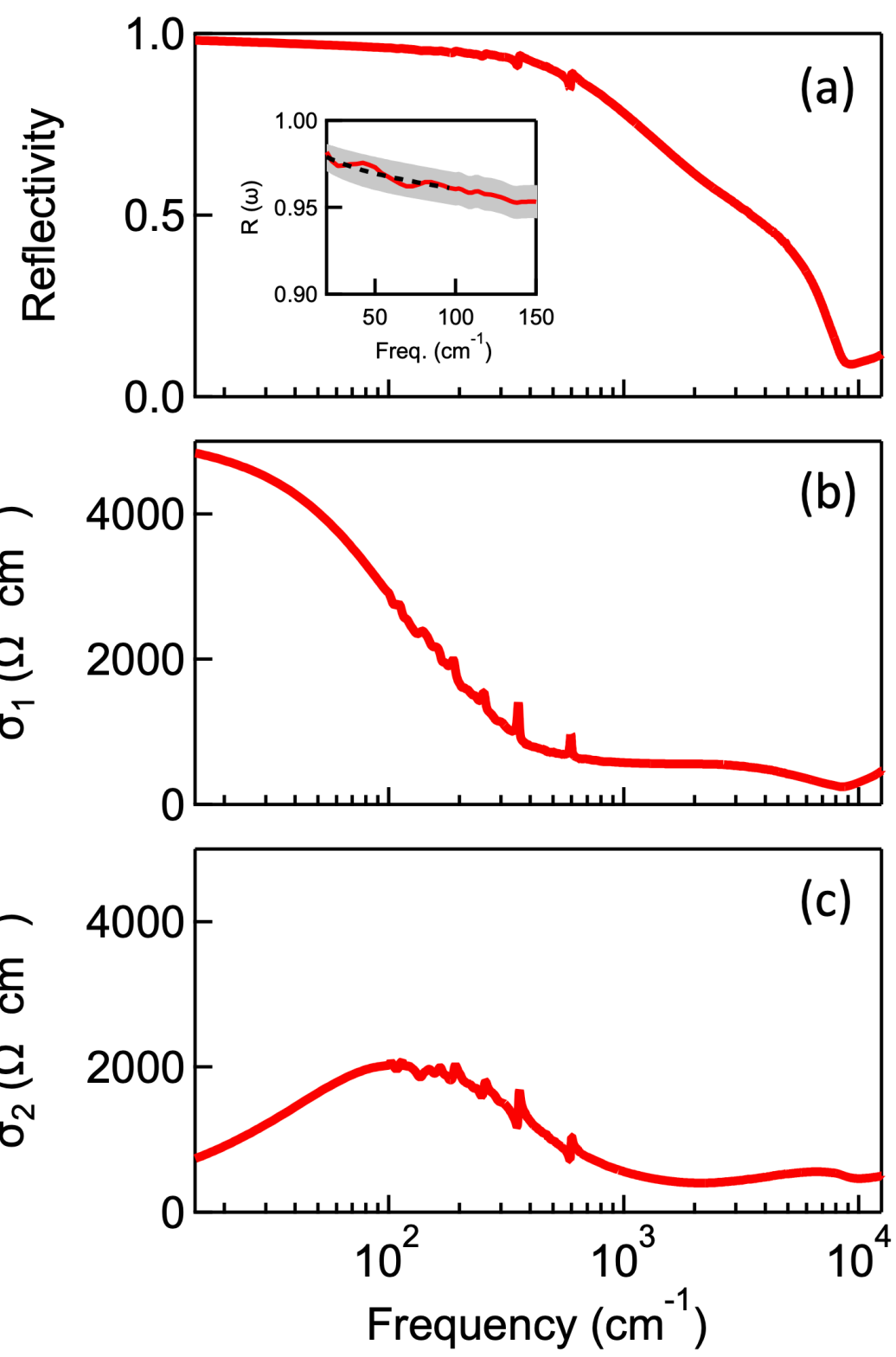

**Figure S2.** (a) Reflectivity and (b,c) complex optical conductivity, $\sigma_1(\omega) + i\sigma_2(\omega)$, determined at equilibrium for $YBa_2Cu_3O_{6.48}$ at $T = 100$ K via Fourier-Transform Infrared Spectroscopy and subsequent Kramers-Kronig analysis. The inset in (a) shows raw reflectivity data down to the lowest measured frequency (red), along with the Drude fit used in the Kramers-Kronig analysis (black dashed). The gray band represents a putative $\pm 1\%$ uncertainty bar in the reflectivity level (see Fig. S7 and related discussion).

## S3. Pump-probe setup and data acquisition

To retrieve the photoinduced changes in the optical properties of $YBa_2Cu_3O_{6.48}$ we performed mid-infrared pump / THz probe experiments with different pump pulse durations (see Fig. S3). The $YBa_2Cu_3O_{6.48}$ crystal was manually polished in order to expose a surface which contained both the *a* (in-plane) and *c* (out-of-plane) axis.

The sample was then glued on top of a cutting-edge-shaped holder to prevent stray reflections from interfering with the signal from the sample surface. This holder was installed on the cold finger of a He-flow cryostat to allow for temperature control.

$YBa_2Cu_3O_{6.48}$ was photoexcited with mid-infrared pump pulses of variable duration between ~300 fs and ~2.5 ps, tuned to ~20 THz center frequency (~15 μm wavelength), in resonance to the $B_{1u}$ phonon modes that were previously shown to enhance interlayer superconducting coupling.

These pump pulses were generated by difference frequency mixing of the signal and idler outputs of an optical parametric amplifier (OPA) in a 0.35 mm thick GaSe crystal which yielded 15-μm-center-wavelength pulses with total energy of ~15 μJ and typical duration of 300 fs. The OPA was pumped with ~80 fs long pulses from a commercial Ti:sapphire regenerative amplifier (800 nm wavelength). The duration of the pump pulses was controlled by introducing dispersion through propagation in NaCl rods and their intensity was adjusted using two freestanding wire-grid polarizers.

The *p*-polarized pump pulses were then focused onto the sample surface, with their electric field aligned with the *c* axis of the $YBa_2Cu_3O_{6.48}$ crystal. The typical pump spot size was ~0.5 mm allowing for a homogenous illumination of the probed area. A maximum fluence of ~8 $mJ/cm^2$ could be achieved.

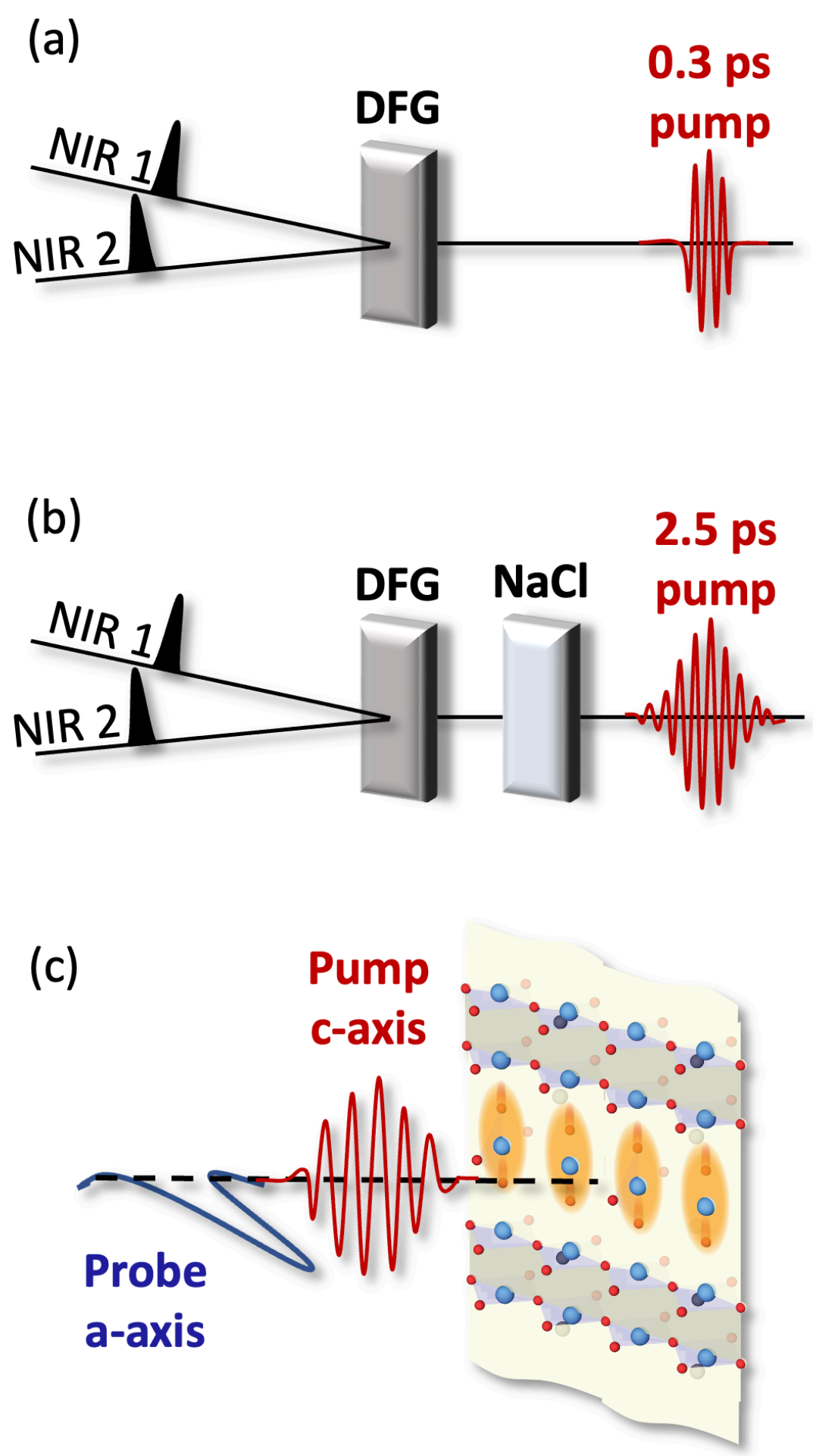

**Figure S3.** Pump-probe experimental setup. (a) The signal and idler outputs from an optical parametric amplifier are sent to a GaSe crystal for difference frequency generation (DFG). The duration of these pulses can be controlled by letting them propagate through optically polished NaCl plates (b). (c) The mid-infrared pump pulses (red) are polarized along the *c* axis, while the single-cycle THz probe (blue) pulses along the *a* axis of $YBa_2Cu_3O_{6.48}$.

The transient reflectivity changes after photoexcitation were retrieved via time-domain THz spectroscopy using single-cycle THz pulses generated by illuminating a 1-mm-thick ZnTe (110) crystal with 800 nm pulses. These *s*-polarized probe pulses, with their electric field oriented along the *a* crystallographic axis of $YBa_2Cu_3O_{6.48}$, were focused at normal incidence onto the sample surface on a spot of ~0.3 mm diameter. After reflection from the sample surface, their electric field was detected via electro-optic sampling in a 500-μm thick optically contacted ZnTe crystal. This setup provided a spectral bandwidth extending from ~15 to 80 cm$^{-1}$.

In order to minimize the effects on the pump-probe time resolution due to the finite duration of the THz probe pulse, we performed the experiment as described in Ref. [iv]. The transient reflected field at each time delay $\tau$ after excitation was obtained by keeping fixed the delay $\tau$ between the pump pulse and the electro-optic sampling gate pulse, while scanning the delay $t$ of the single-cycle THz probe pulse. The stationary probe electric field $E_R(t)$ and the differential electric field $\Delta E_R(t,\tau)$ reflected from the sample were recorded simultaneously by feeding the electro-optic sampling signal into a digitizer that sampled individual laser pulses, and mechanically chopping the pump and probe beams at different frequencies. $E_R(t)$ and $\Delta E_R(t,\tau)$ were then independently Fourier transformed to obtain the complex-valued, frequency-dependent $\tilde{E}_R(\omega)$ and $\Delta\tilde{E}_R(\omega,\tau)$.

Note that, for comparison with the experiments reported earlier, we also preformed measurements with *p*-polarized THz probe pulses, *i.e.*, with the electric field oriented along the $c$ (out-of-plane) direction. We obtained virtually identical results to those reported in Ref. [v].

## S4. Determination of the transient optical properties

From the frequency-dependent measured quantities $\tilde{E}_R(\omega)$ and $\Delta\tilde{E}_R(\omega,\tau)$ the complex reflection coefficient in the photoexcited state, $\tilde{r}(\omega,\tau)$, could be extracted by inverting the following equation:

$$\frac{\Delta\tilde{E}_R(\omega,\tau)}{\tilde{E}_R(\omega)} = \frac{\tilde{r}(\omega,\tau)-\tilde{r}_0(\omega)}{\tilde{r}_0(\omega)} \qquad (1)$$

As the penetration depth ($\approx 1$ μm) of the excitation pulses (defined as $d(\omega) = \frac{c}{2\omega} Im[\tilde{n}_0(\omega)]$, where $\tilde{n}_0(\omega)$ is the equilibrium complex refractive index) was typically larger than that of the probe pulses ($\approx 0.2 - 0.7$ μm), the THz probe pulse sampled a homogeneously excited volume. The transient optical properties could then be extracted directly, without the need to consider any pump-probe penetration depth mismatch in the data analysis.

In this case, the complex refractive index of the photo-excited material, $\tilde{n}(\omega, \tau)$, was directly retrieved from the Fresnel relation:

$$\tilde{n}(\omega, \tau) = \frac{1-\tilde{r}(\omega,\tau)}{1+\tilde{r}(\omega,\tau)} \quad (2)$$

and from this, the transient complex optical conductivity, $\tilde{\sigma}(\omega, \tau) = \frac{\omega}{4\pi i}[\tilde{n}(\omega, \tau)^2 - \varepsilon_\infty]$.

Here, we took $\varepsilon_\infty = 4.5$, a standard value for high-$T_C$ cuprates [vi].

In Fig. S4 we report an example of raw pump-induced changes in the complex reflection coefficient, $\frac{\tilde{r}(\omega,\tau)-\tilde{r}_0(\omega)}{\tilde{r}_0(\omega)}$. Corresponding reflectivities in equilibrium and after photo-excitation are also shown.

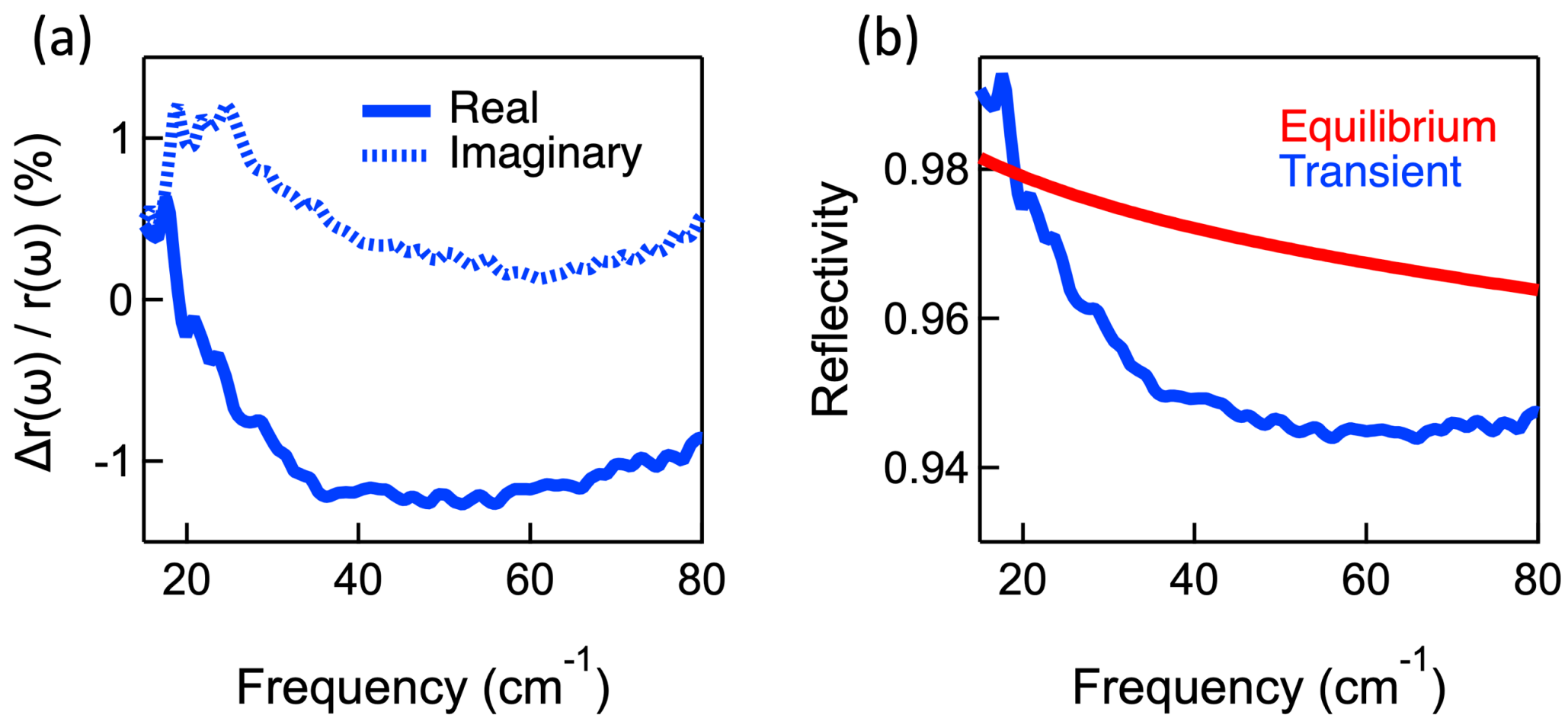

**Figure S4.** (a) Raw pump-induced changes in the complex reflection coefficient, $\frac{\tilde{r}(\omega,\tau)-\tilde{r}_0(\omega)}{\tilde{r}_0(\omega)}$, measured in $YBa_2Cu_3O_{6.48}$ at $T = 100$ K upon excitation with ~0.3 ps long mid-infrared pulses with a pump fluence of ~8 mJ/cm$^2$. Corresponding reflectivity in equilibrium (red) and after photo-excitation (blue), calculated as $R_0(\omega) = |\tilde{r}_0(\omega)|^2$ and $R(\omega, \tau) = |\tilde{r}(\omega, \tau)|^2$, respectively (no modeling for penetration depth mismatch involved).

For accurate comparison with data acquired previously with probe pulses polarized along the $c$ (out-of-plane) axis, a configuration for which the penetration depth of the pump was smaller than that of the probe, we also performed the same type of analysis reported in Refs. [v,vii].

We modeled a photoexcitation condition that is not completely homogeneous by "slicing" the probed thickness of the material into thin layers, where we assumed that the pump-induced changes in the refractive index were either proportional to the pump intensity in the layer, *i.e.*, $\tilde{n}(\omega, z) = \tilde{n}_0(\omega) + \Delta\tilde{n}(\omega)e^{-z/d(\omega)}$, or to the pump electric field, *i.e.*, $\tilde{n}(\omega, z) = \tilde{n}_0(\omega) + \Delta\tilde{n}(\omega)e^{-z/[2d(\omega)]}$. Here, $z$ is the spatial coordinate along the sample thickness. In addition, we also considered a saturation profile as that introduced in Refs. [viii,ix], making use of the fluence dependence reported in Fig. S5.

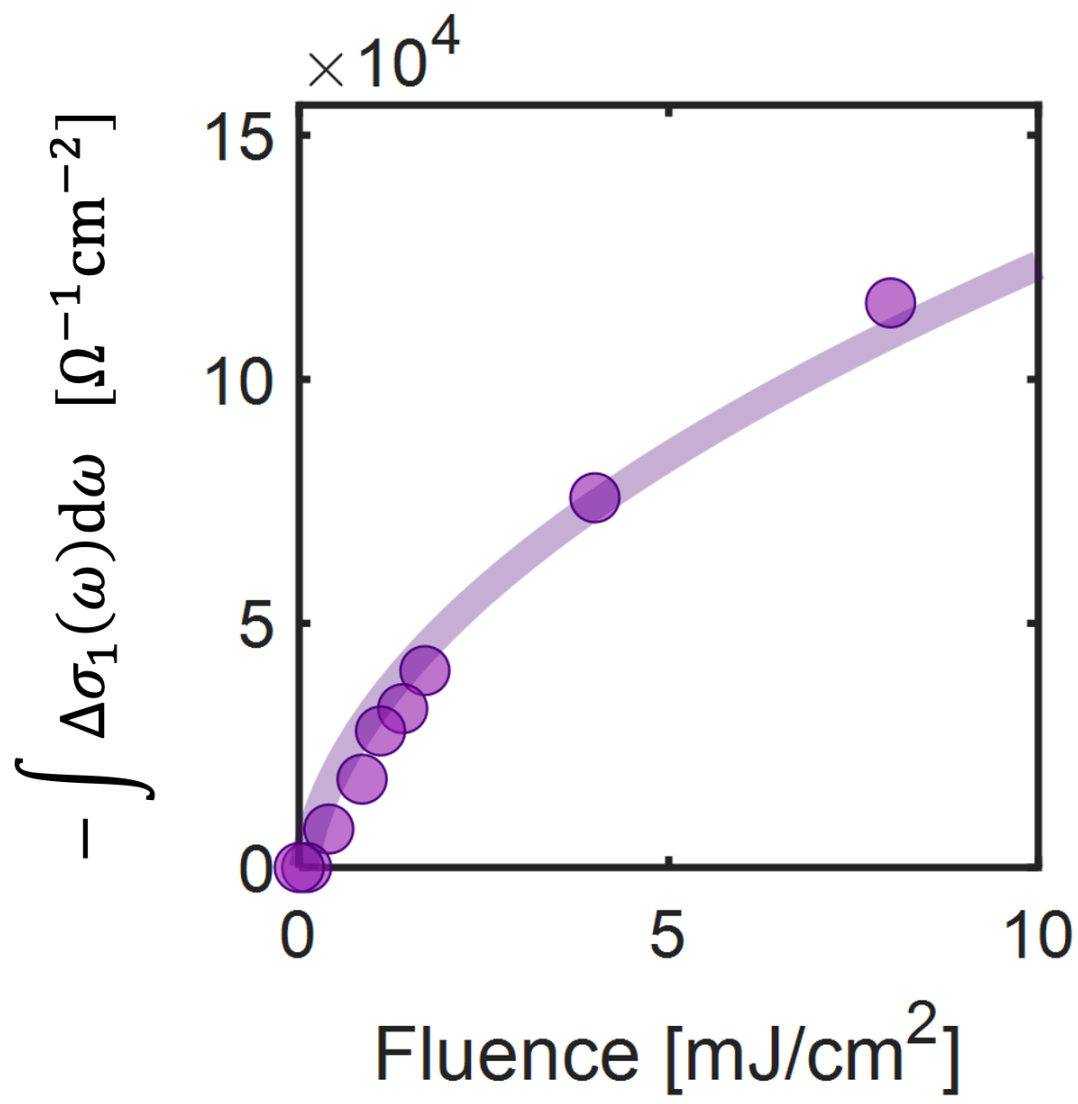

**Figure S5.** Fluence dependence of the pump-induced frequency-integrated changes in $\sigma_1(\omega)$, measured in $YBa_2Cu_3O_{6.48}$ at $T = 100$ K for ~0.3 ps long mid-infrared excitation pulses. The line is a square-root fit.

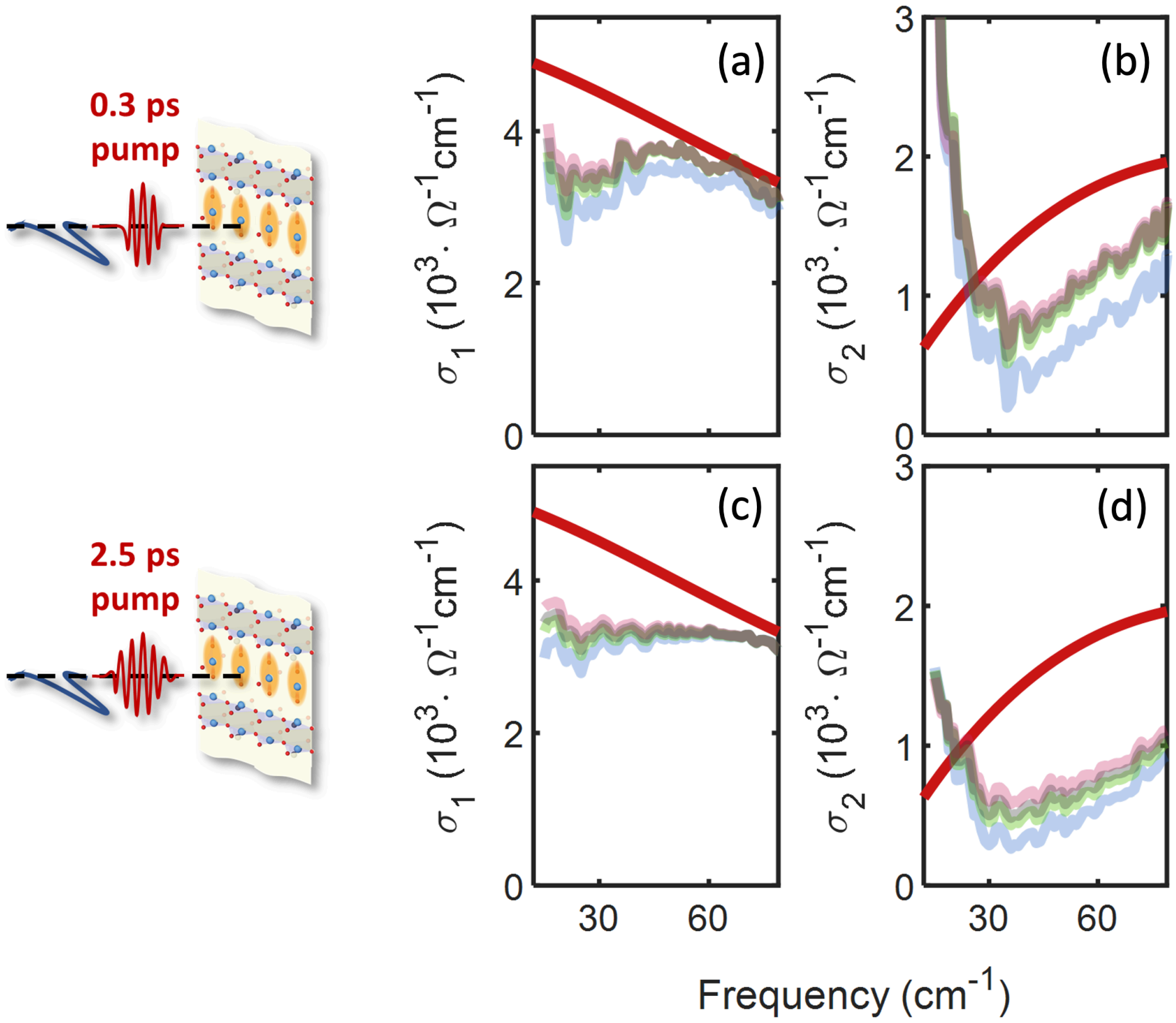

**Figure S6.** Complex $a$-axis optical conductivity, $\sigma_1(\omega) + i\sigma_2(\omega)$, measured in $YBa_2Cu_3O_{6.48}$ at $T =$ 100 K in equilibrium (red) and at the peak of the photoinduced response after excitation with two different mid-infrared pulse durations (shown on the left) and for the same pump fluence of ~6.5 mJ/cm$^2$. We compare spectra extracted directly with the Fresnel equations for a semi-infinite medium (no penetration depth mismatch, purple), with those calculated with a multilayer model which assumed pump-induced changes proportional to the pump electric field (green) or to the pump intensity (blue) in each layer. Spectra calculated with the saturation model introduced in Refs. [viii,ix] are also shown (gray). Here, the fluence dependence reported in Fig. S5 has been used to extract the saturation profile for $\Delta\tilde{n}(\omega)$.

For each probe frequency $\omega$, the complex reflection coefficient $\tilde{r}(\Delta\tilde{n})$ of such a multilayer stack was calculated with a characteristic matrix approach[x], keeping $\Delta\tilde{n}$ as a free parameter. Equation (1) relates the measured quantity $\frac{\Delta\tilde{E}_R(\omega,\tau)}{\tilde{E}_R(\omega)}$ to the changes in the complex reflection coefficient. One can extract $\Delta\tilde{n}(\omega)$ by minimizing numerically:

$$\left|\frac{\Delta\tilde{E}_R(\omega,\tau)}{\tilde{E}_R(\omega)} - \frac{\tilde{r}(\omega,\Delta\tilde{n})-\tilde{r}_0(\omega)}{\tilde{r}_0(\omega)}\right| \qquad (3)$$

The value of $\Delta\tilde{n}(\omega)$ retrieved by this minimization represents the pump-induced change in the refractive index at the surface. By taking $\tilde{n}(\omega) = \tilde{n}_0(\omega) + \Delta\tilde{n}(\omega)$, one can reconstruct the optical response functions of the material as if it had been homogeneously transformed by the pump.

Fig. S6 shows a comparison between the in-plane perturbed optical properties of $YBa_2Cu_3O_{6.48}$ extracted using the different analysis approaches discussed above. As can be clearly seen, the variations are negligible.

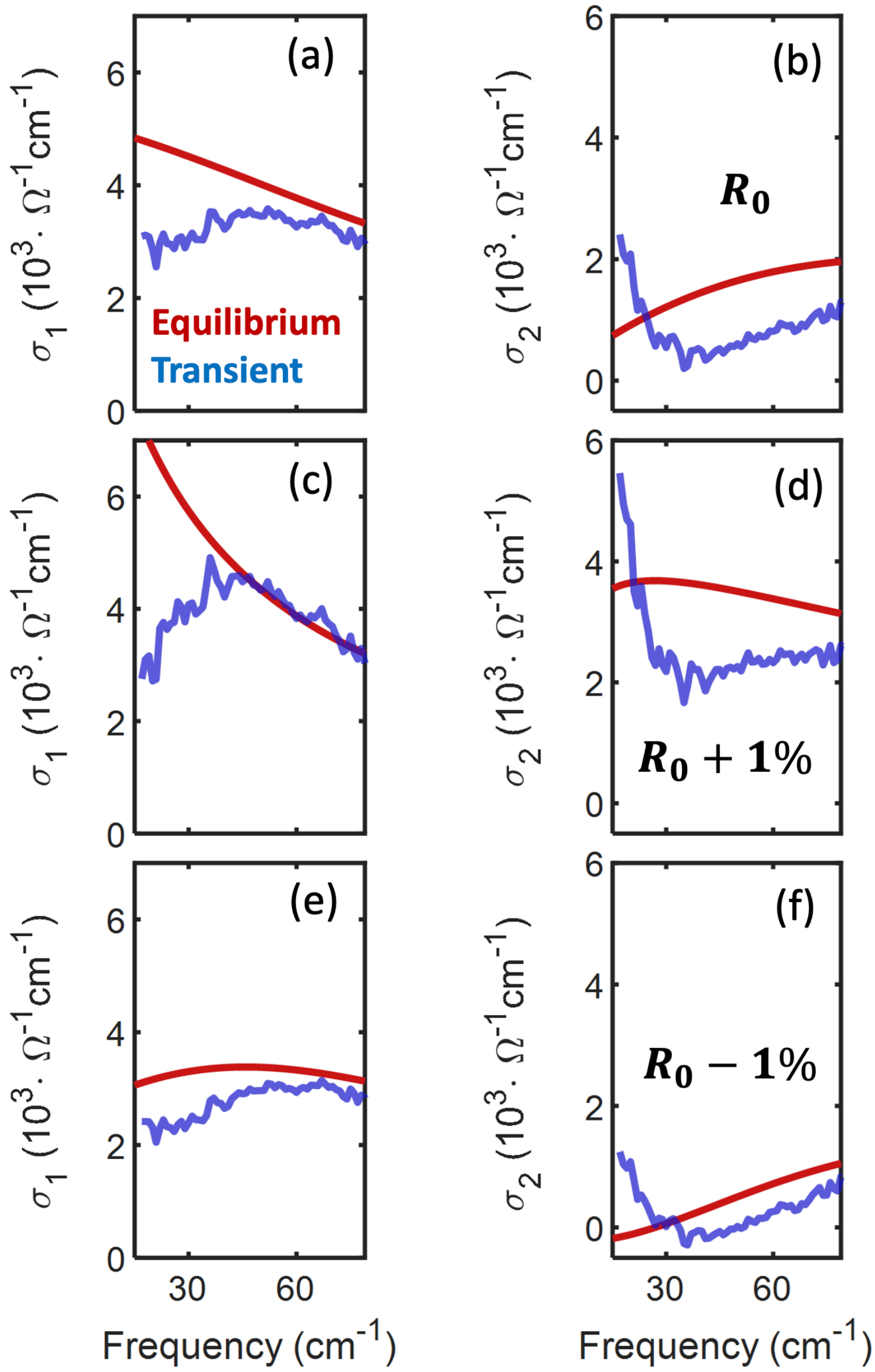

**Figure S7.** Complex $a$-axis optical conductivity, $\sigma_1(\omega) + i\sigma_2(\omega)$, measured in $YBa_2Cu_3O_{6.48}$ at $T =$ 100 K in equilibrium (red) and at the peak of the photoinduced response after excitation with ~0.3 long mid-infrared pulses with a same pump fluence of ~6.5 mJ/cm$^2$ (blue). We show spectra derived from the original equilibrium reflectivity $R_0$ (a,b), as well as spectra calculated assuming an equilibrium reflectivity 1% higher (c,d) or 1% lower (e,f).

Another possible source of uncertainty in the determination of the transient optical properties lies in the extraction procedure of the equilibrium complex conductivity. In Fig. S7 we show an analysis of how a realistic $\pm 1\%$ uncertainty in the equilibrium reflectivity would affect the extracted complex optical conductivities. Also in this case, the "superconducting-like" character of the photoinduced response, *i.e.*, partial $\sigma_1(\omega)$ gapping and divergence in $\sigma_2(\omega)$, is not affected.

## S5. Fitting models, extracted parameters, and extended data sets

The in-plane equilibrium optical response of $YBa_2Cu_3O_{6.48}$ was fitted with a Drude-Lorentz model, for which the complex conductivity is expressed as:

$$\tilde{\sigma}_a(\omega) = \frac{\omega_p^2}{4\pi}\frac{1}{\Gamma - i\omega} + \tilde{\sigma}_{HF}(\omega) \qquad (4)$$

Here, $\omega_p$ and $\Gamma$ are the Drude plasma frequency and momentum relaxation rate, while $\tilde{\sigma}_{HF}(\omega) = \sum_i \frac{\Omega_{p,i}^2}{4\pi}\frac{\omega}{i(\Omega_{0,i}^2 - \omega^2) + \gamma_i\omega}$ is a sum of Lorentz oscillators with peak frequency $\Omega_{0,i}$ and, plasma frequency $\Omega_{p,i}$, and damping coefficient $\gamma_i$. These high-frequency terms, associated with both electronic absorptions and phonon modes (only partially shielded by free charges in the metal), were determined by fitting the broadband complex conductivity measured with Fourier transform infrared spectroscopy and kept fixed for the whole analysis reported in this paper.

The same Drude model of Eq. (4) was also employed to fit the transient optical spectra after photo-excitation, by letting only $\omega_p$ and $\Gamma$ vary from their equilibrium values. Importantly, for each data set, the same parameters were used to simultaneously fit the reflectivity, the real part of the optical conductivity, as well as its imaginary part.

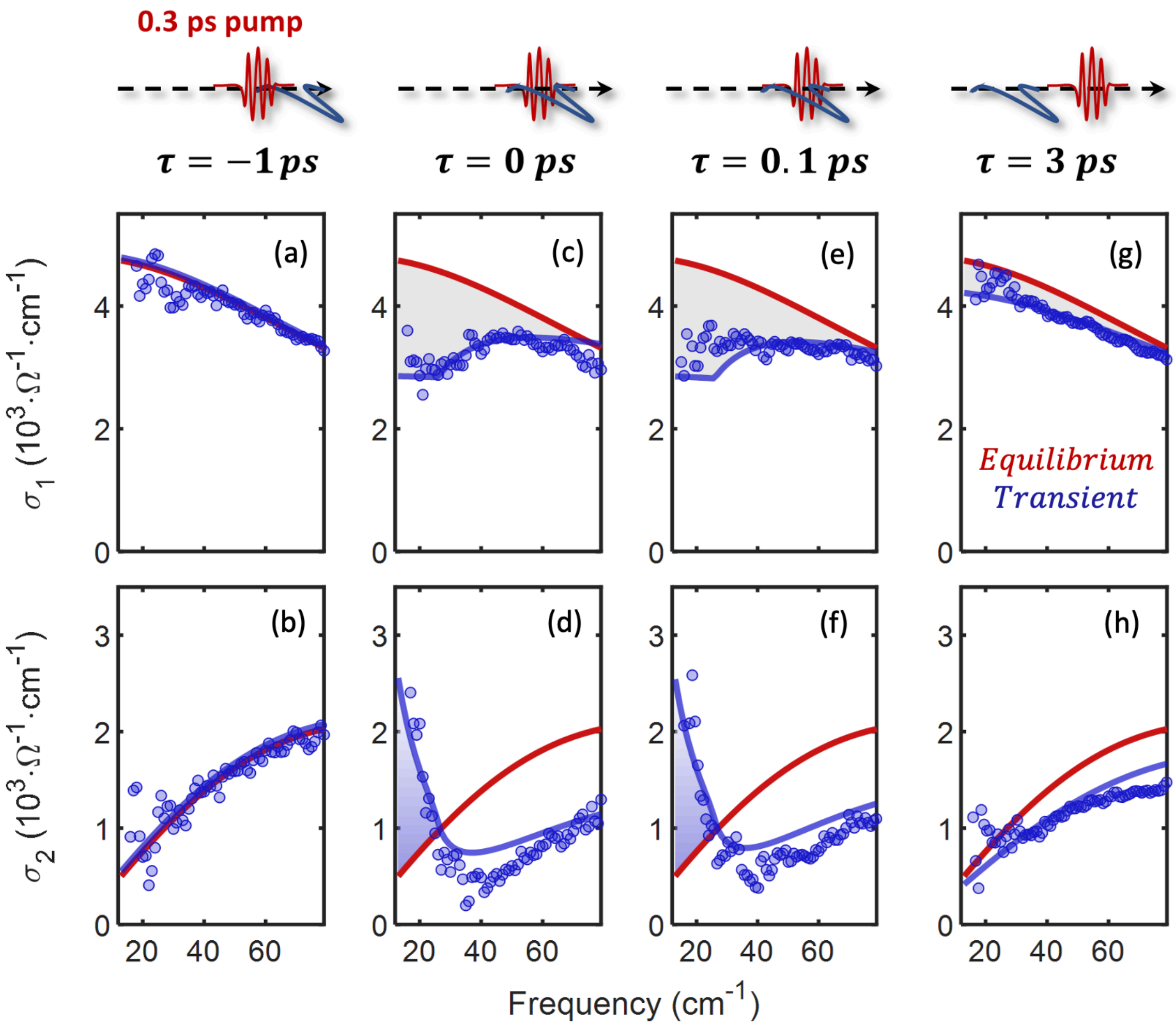

**Figure S8.** Complex $a$-axis optical conductivity, $\sigma_1(\omega) + i\sigma_2(\omega)$, measured in $YBa_2Cu_3O_{6.48}$ at $T = 100$ K in equilibrium (red) and at different selected pump-probe time delay throughout the photoinduced dynamics (blue circles). We display data taken for a mid-infrared pulse duration of 0.3 ps and a pump fluence of ~6.5 mJ/cm$^2$. Blue lines are fits to the spectra with the two fluid model described in the text.

Only for time delays and excitation fluences for which a superconducting-like response was found, it was necessary to extend our fitting model. In this case, following the approach taken in Ref. [v] for the out-of-plane response, we used a two-fluid model. The first fluid describes the response of a superconductor with the Zimmermann model [xi], *i.e.*, an extension of the Mattis-Bardeen model for superconductors of arbitrary purity, having the optical gap, 2Δ, as the fit parameter. The second fluid consists instead of unpaired residual quasiparticles[xii], whose response is described by a Drude model with increased scattering rate, Γ, with respect to the equilibrium value, in order to account for carrier heating. In addition, to preserve the total number of carriers we introduced a filling fraction parameter, $f$, so that the total fit formula is expressed as:

$$\tilde{\sigma}_a(\omega) = f\tilde{\sigma}_{Zim}(\omega, 2\Delta) + (1-f)\frac{\omega_P^2}{4\pi}\frac{1}{\Gamma - i\omega} + \tilde{\sigma}_{HF}(\omega) \quad (5)$$

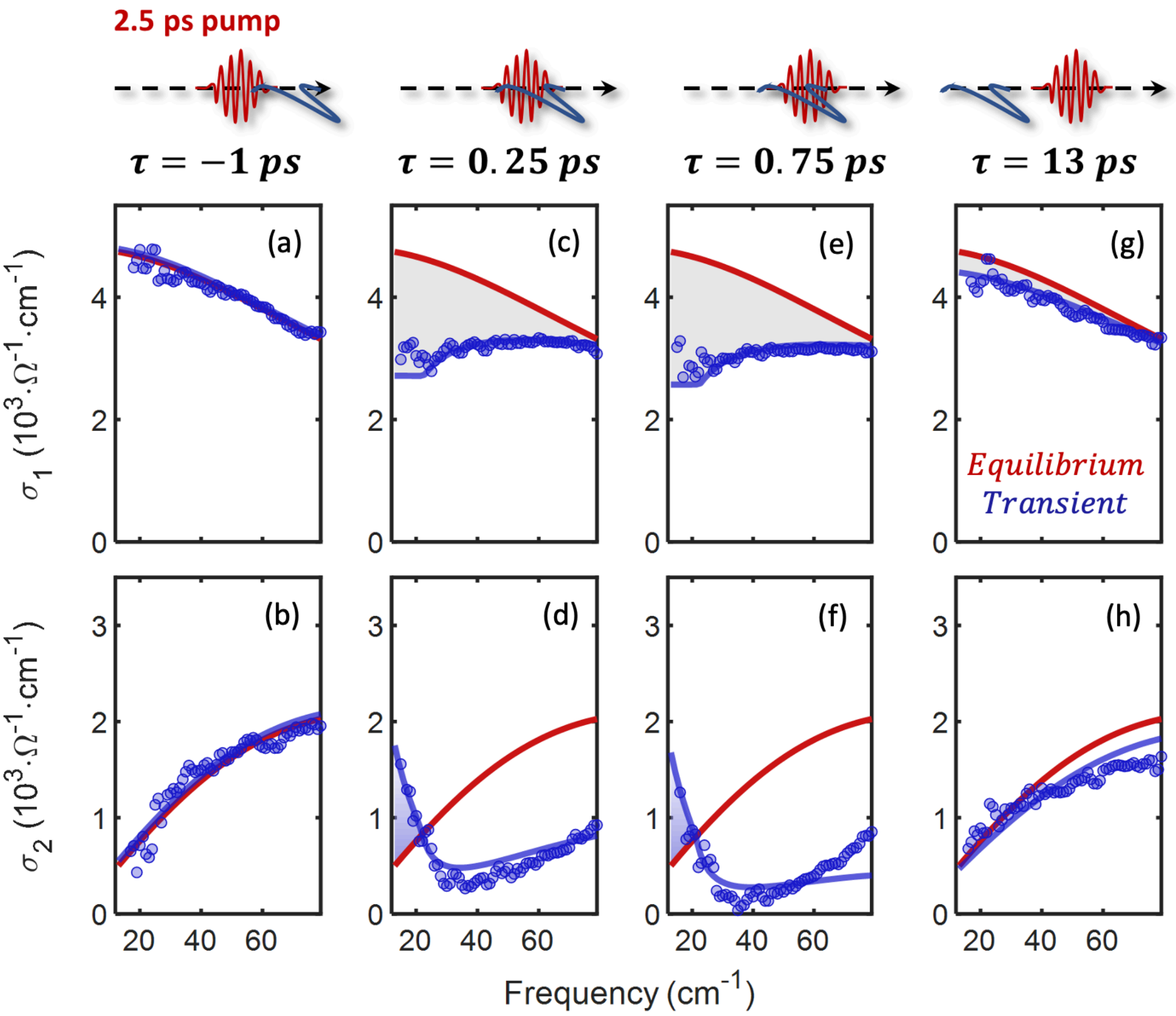

**Figure S9.** Complex $a$-axis optical conductivity, $\sigma_1(\omega) + i\sigma_2(\omega)$, measured in $YBa_2Cu_3O_{6.48}$ at $T = 100$ K in equilibrium (red) and at different selected pump-probe time delay throughout the photoinduced dynamics (blue circles). We display data taken for a mid-infrared pulse duration of 2.5 ps and a pump fluence of ~6.5 mJ/cm². Blue lines are fits to the spectra with the two fluid model described in the text.

Examples of fits are given in Fig. S8 and Fig. S9, where we report extended data sets taken for two different pump pulse durations and different time delays throughout the photoinduced dynamics. In both figures it can be seen that the response develops a prompt reduction in the real part of the conductivity, $\sigma_1(\omega)$, and a positive low-frequency divergence in the imaginary part, $\sigma_2(\omega)$, as soon as the pump hits the sample. These then relax on different time scales depending on the duration of the excitation pulse until the equilibrium properties are recovered at long positive time delays.

We stress here that the divergence found in $\sigma_2(\omega)$ at low frequencies at the peak of the response is entirely enclosed in the superconducting term, $\tilde{\sigma}_{Zim}(\omega, 2\Delta)$, while the reduction in $\sigma_1(\omega)$ is partly to be attributed to quasiparticles heating, which appears as a rigid downward shift and broadening, and partly to the opening of the gap in $\tilde{\sigma}_{Zim}$.

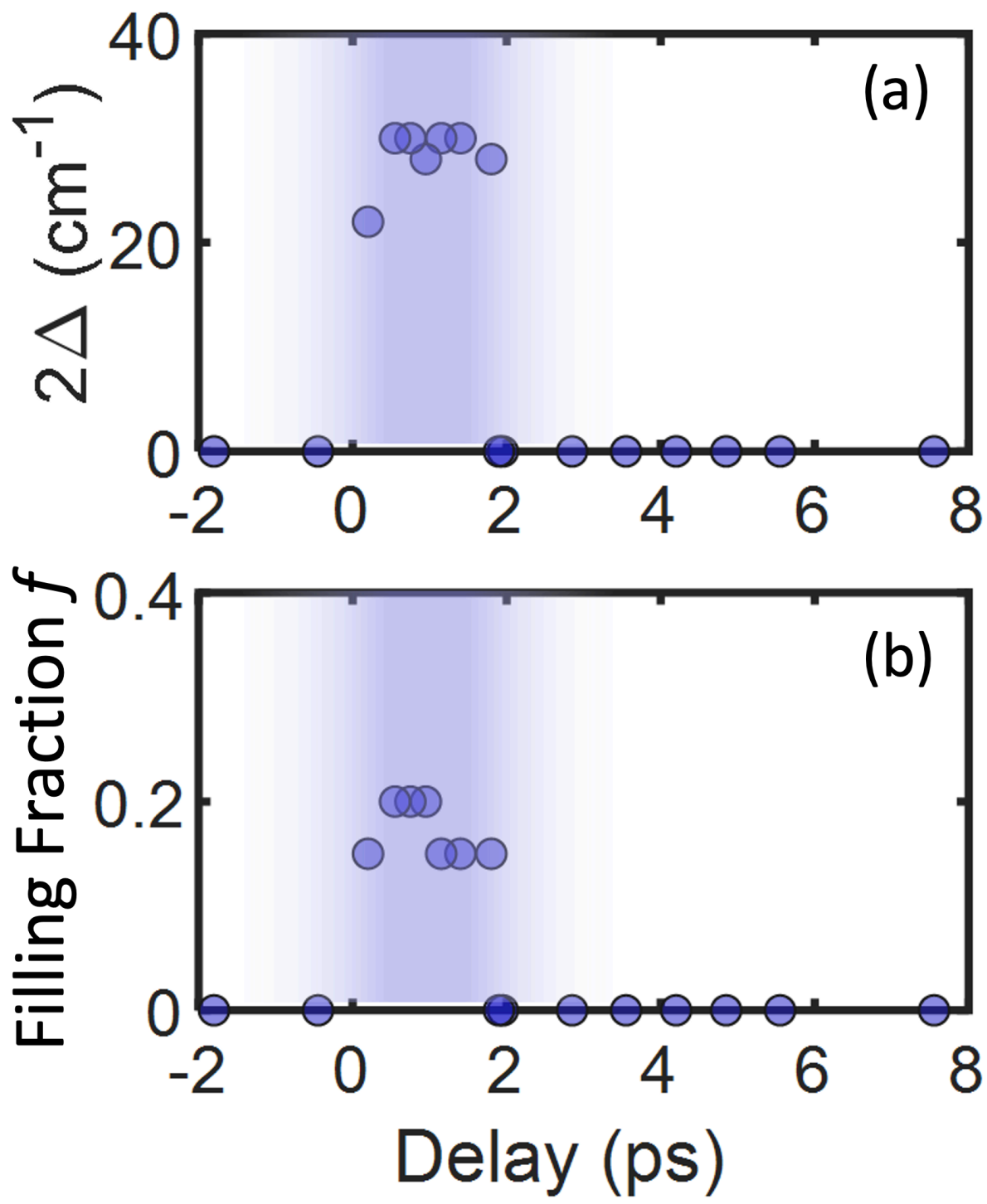

**Figure S10.** Dynamical evolution of the optical gap, 2Δ, and the superconducting filling fraction, $f$, as a function of pump-probe time delay, extracted from two-fluid fits (see main text) to transient $\sigma_1(\omega) + i\sigma_2(\omega)$ measured in $YBa_2Cu_3O_{6.48}$ at $T = 100$ K upon excitation with 2.5 ps long pulses. The shaded areas represent the time delay window for which a finite coherent, superconducting-like response was detected.

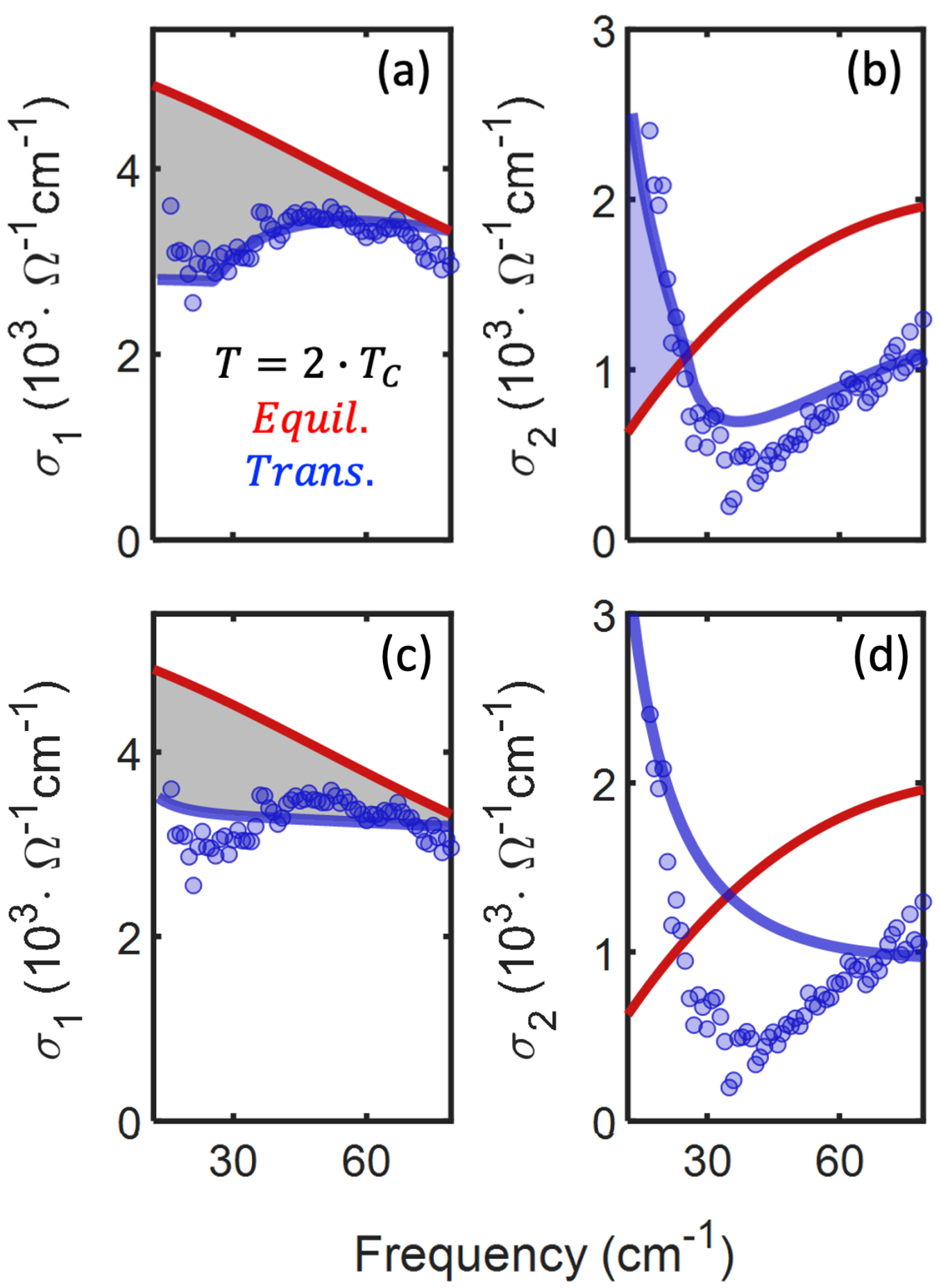

**Figure S11.** Complex $a$-axis optical conductivity, $\sigma_1(\omega) + i\sigma_2(\omega)$, measured in $YBa_2Cu_3O_{6.48}$ at $T = 100$ K in equilibrium (red) at the peak of the photoinduced response after excitation with ~0.3 ps long mid-infrared pulses with a pump fluence of ~6.5 mJ/cm$^2$ (blue circles). Blue lines in (a,b) are fits to the spectra with the two fluid model described in the text. Those in (c,d) are attempted fits with a model with two Drude terms, one of which has a scattering rate of a few cm$^{-1}$, *i.e.*, far below the measurement range.

The extracted parameters for the Drude term at equilibrium are $\omega_p \simeq 6000$ cm$^{-1}$ and $\Gamma \simeq 150$ cm$^{-1}$, while at the peak of the photoinduced response we obtained typical values of $\Gamma \simeq 400$ cm$^{-1}$, $2\Delta \simeq 30$ cm$^{-1}$, and $f \simeq 0.2$. In Fig. S10 we report the dynamical evolution of the superconducting parameters, namely $2\Delta$ and $f$, for the dataset acquired with ~2.5 ps long pump pulses.

The need for a two-fluid model, one of which has superconducting properties, to describe the response at short delay times is apparent from the analysis reported in Fig. S11. Therein, we show a comparison applied to a specific dataset between fits with the two-fluid model introduced above and attempted fits with a model with two Drude terms, one of which has a finite scattering rate of a few $\mathrm{cm}^{-1}$, *i.e.*, far below the measurement range. Although the latter model is at least partially able to reproduce a low-frequency $\sigma_2(\omega)$ divergence, it is only the two-fluid model that can capture even quantitatively all experimentally observed spectral features, including the partial gapping in $\sigma_1(\omega)$.

We now turn to the optical determination of a quantity that in a superconductor is proportional to the superfluid density. As shown in Fig. 4(d) of the main text, this can be extracted in a model-independent way by estimating the strength of the low-frequency divergence in $\sigma_2(\omega)$. To this end, we systematically fitted all measured spectra for which a $\sigma_2(\omega)$ divergence was observed with the formula $\sigma_2(\omega) \propto 1/\omega$. These fits, an example of which is shown in Fig. 12(a), were limited to the frequency range $\omega \lesssim 30\ \mathrm{cm}^{-1}$. The error bars, expressed by the standard deviation, are reported for a full data set in Fig. S12(b).

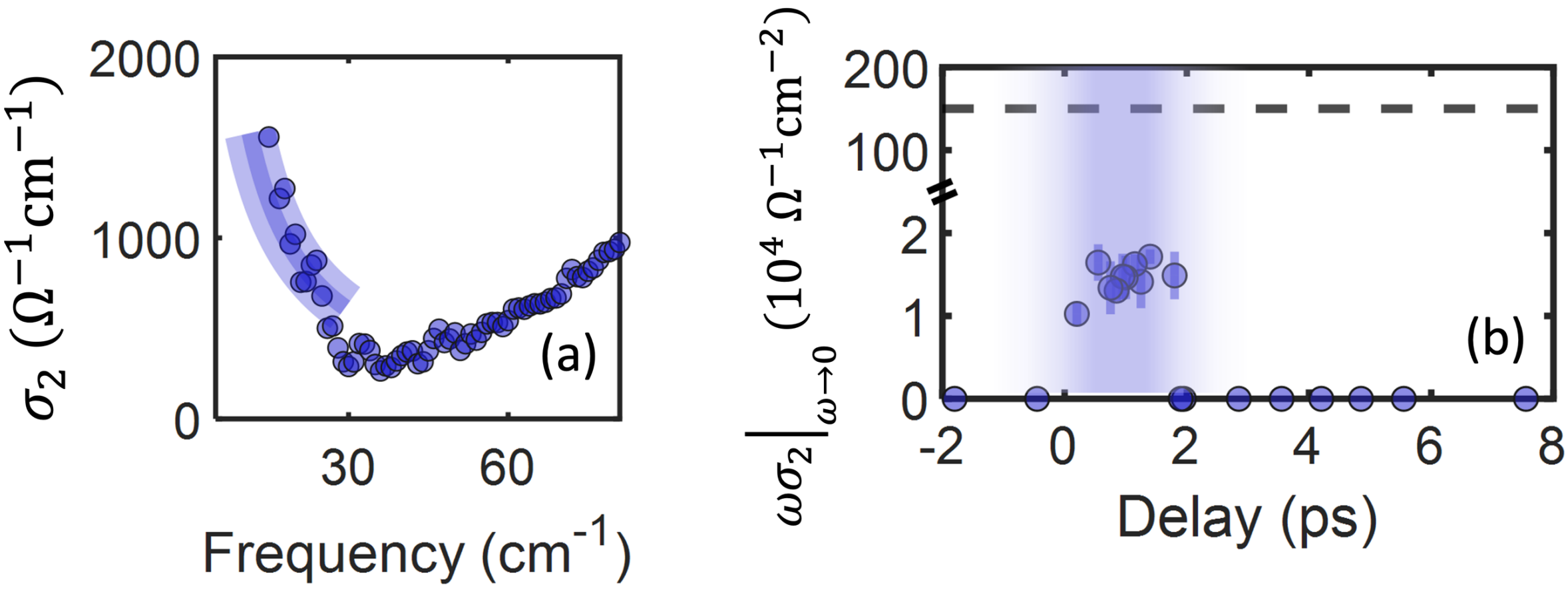

**Figure S12.** (a) Imaginary part of the $a$-axis conductivity, $\sigma_2(\omega)$, measured in $YBa_2Cu_3O_{6.48}$ at $T = 100$ K in equilibrium after excitation with ~2.5 ps long mid-infrared pulses with a pump fluence of ~6.5 $\mathrm{mJ/cm^2}$ (blue circles). The line is a $\propto 1/\omega$ fit, while the light-blue band indicates the standard deviation in the least square fitting procedure. (b) Dynamical evolution of $\lim_{\omega\to 0} \omega\sigma_2(\omega)$, where the error bars come from the standard deviation of the fits in (a). For all time delays where no low-frequency divergence was found in $\sigma_2(\omega)$, the value of $\lim_{\omega\to 0} \omega\sigma_2(\omega)$ was set to be zero.

A useful cross check to verify the validity of this procedure makes use of the relation:

$$\int_0^{\Omega}[\sigma_{1,N}(\omega) - \sigma_{1,S}(\omega)]\, \mathrm{d}\omega = \frac{\pi}{2} \lim_{\omega\to 0}[\omega\sigma_{2,S}(\omega)] \quad (5)$$

where $\sigma_N$ and $\sigma_S$ are the normal-state and superconducting-state conductivities, respectively. Equation (5) is valid for any superconductor as long as the range of integration in the left term starts at zero frequency and extends to sufficiently high frequencies to capture all condensed spectral weight.

Having no data available down to zero frequency we made use of a model introduced in Ref. [xiii] (see also https://reffit.ch/wp-content/uploads/2018/10/Manual.pdf) whereby the real part of the conductivity of a "gapped Drude" is expressed as:

$$\sigma_1(\omega) = \left\{\xi_0 + (1-\xi_0)\frac{1}{2}\left[1+\tanh\left(\frac{\omega-2\Delta}{2\delta_\Delta}\right)\right]\right\}\frac{\omega_p^2\Gamma}{4\pi(\Gamma^2+\omega^2)} \quad (6)$$

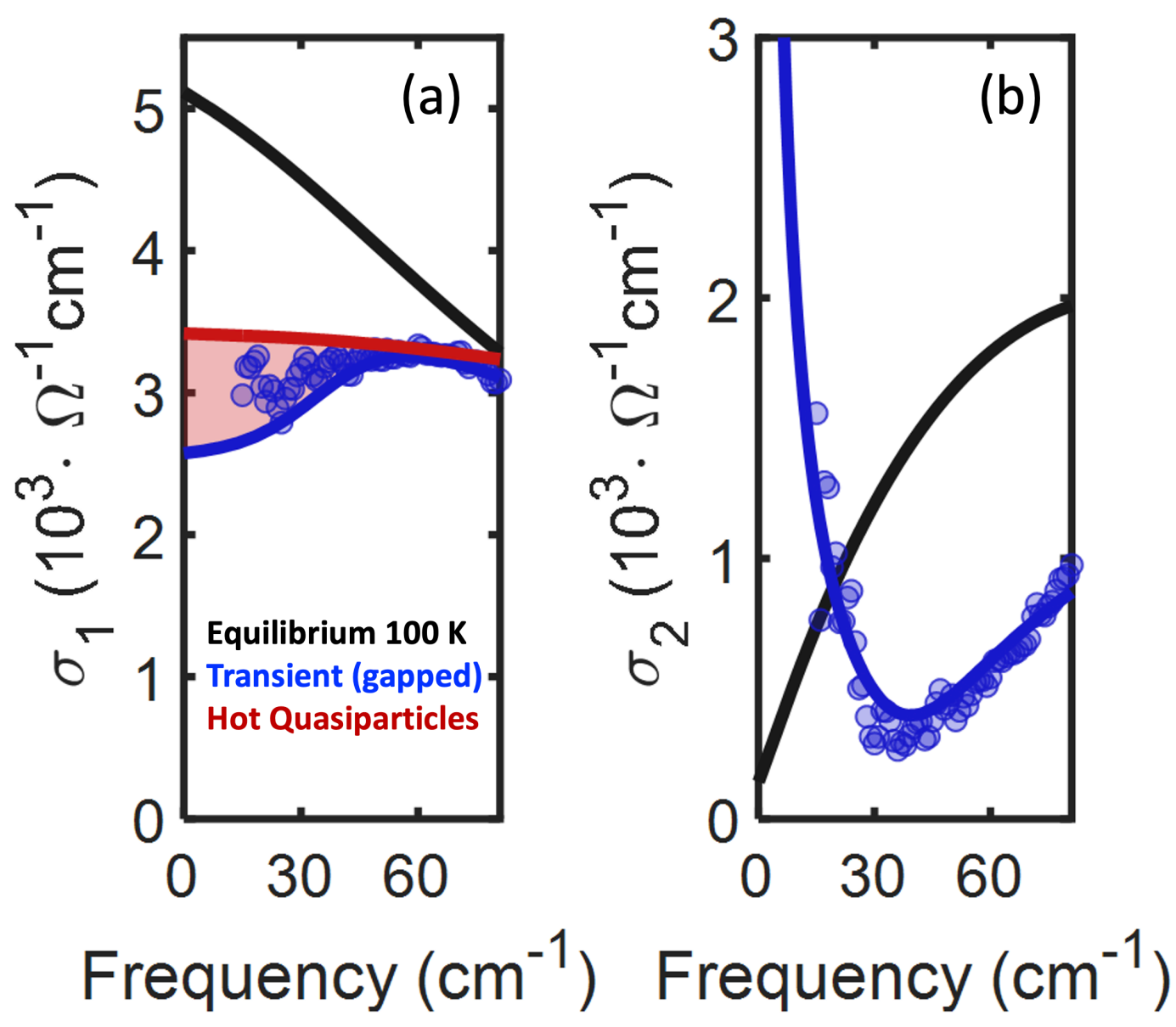

**Figure S13.** Complex $a$-axis optical conductivity, $\sigma_1(\omega) + i\sigma_2(\omega)$, measured in $YBa_2Cu_3O_{6.48}$ at $T = 100$ K in equilibrium (black) at the peak of the photoinduced response after excitation with ~0.3 ps long mid-infrared pulses with a pump fluence of ~6.5 mJ/cm$^2$ (blue circles). Blue lines are fits to the spectra with the "gapped Drude" model introduced in Ref. [xiii] and described in the text. The red curve is the conductivity of hot quasiparticles (extracted from the two-fluid fit), which was used to calculate the integral in Eq. (5).

As shown for a representative data set in Fig. S13, using as fit parameters $2\Delta \simeq 30\ \mathrm{cm}^{-1}$, $\delta_\Delta \simeq 10\ \mathrm{cm}^{-1}$, and $\xi_0 \simeq 0.7$, we were able to reproduce the experimental spectra, thus retrieving a zero-frequency extrapolation for $\sigma_1(\omega)$. This allowed us to obtain estimates for the left term in Eq. (5) as $\int_0^{\Omega=80\ \mathrm{cm}^{-1}} \left[\sigma_{1,N}(\omega) - \sigma_{1,S}(\omega)\right] \mathrm{d}\omega \simeq (2.9 \cdot 10^4 - 9.9 \cdot 10^4)\ \Omega^{-1}\mathrm{cm}^{-2}$, where the higher value was derived using for $\sigma_{1,N}(\omega)$ the black curve in Fig. S13, *i.e.*, the equilibrium conductivity at $T = 100$ K, while for the lower value we used the conductivity of the hot quasiparticles according to the two-fluid model introduced before. We consider the lower estimate ( $2.9 \cdot 10^4\ \Omega^{-1}\mathrm{cm}^{-2}$) as the more accurate one, as it takes into account the response of uncondensed quasiparticles in the transient state. The obtained value shall be compared with the right term in Eq. (5), which for the same data set is $\frac{\pi}{2} \lim_{\omega \to 0} \left[\omega \sigma_{2,S}(\omega)\right] \simeq 4.3 \cdot 10^4\ \Omega^{-1}\mathrm{cm}^{-2}$.

Although the application of this model to zero frequency has clear limitations, given the transient nature of the state, it provides an agreement within ~30-40% between different estimates of the "condensate weight", which shall be taken as further validation for our analysis.

We finally turn to the extraction procedure of the quasiparticle temperature, which is already described at length in the main text. For this purpose, we made use of the two-fluid fit introduced above in a similar way to what was reported in Ref. [v] for the out-of-plane response. By comparing the Drude parameters derived from these fits to the transient spectra with those performed on the equilibrium conductivities measured for different temperatures [i,ii,iii], we could associate a quasi-temperature with each out-of-equilibrium conductivity for different time delays and excitation conditions. This method, along with the extracted calibration curve, is described in Fig. S14.

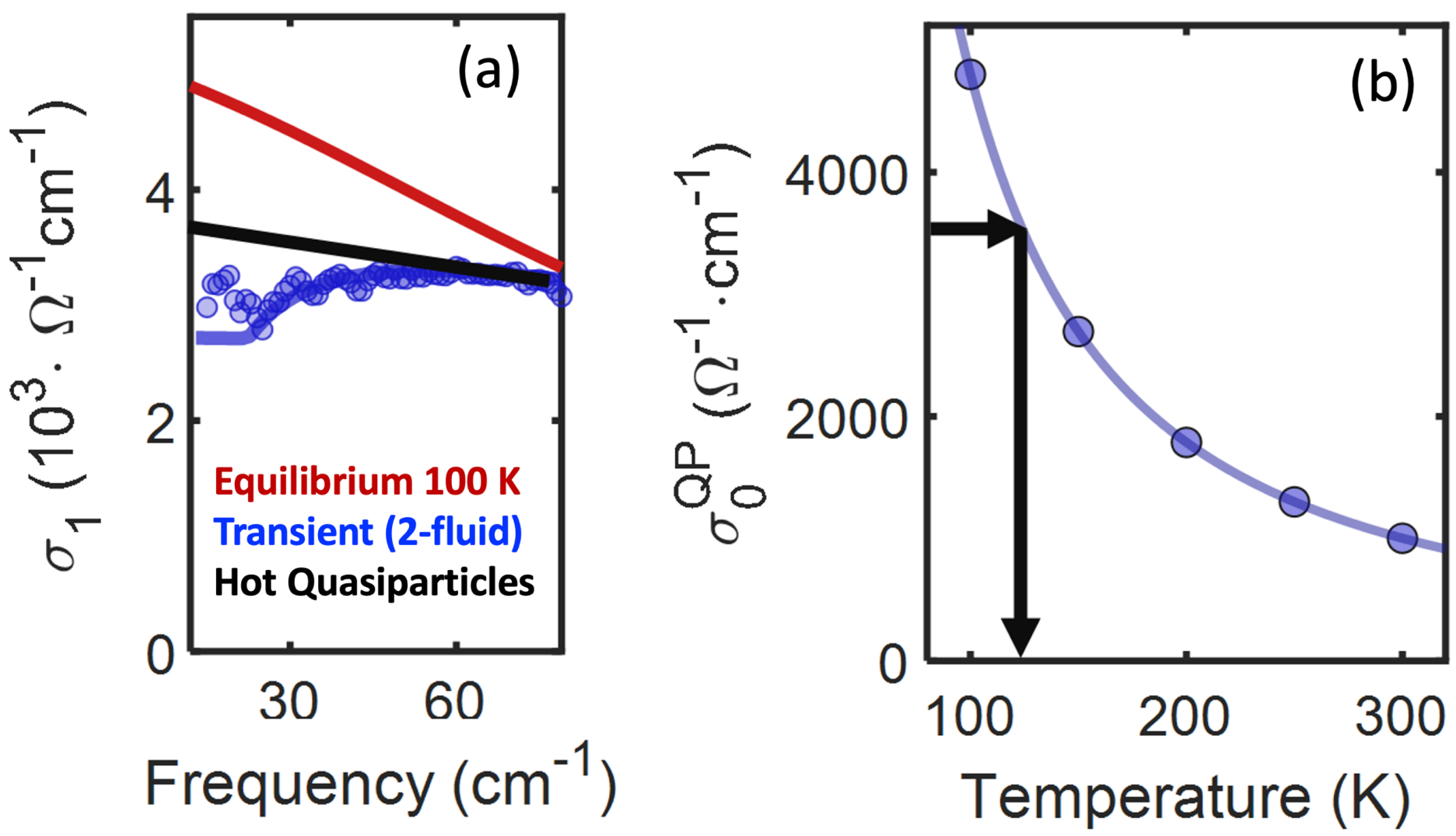

**Figure S14.** (a) Real part of the $a$-axis conductivity, $\sigma_1(\omega)$, measured in $YBa_2Cu_3O_{6.48}$ at $T = 100$ K in equilibrium (red) at the peak of the photoinduced response after excitation with ~2.5 ps long mid-infrared pulses with a pump fluence of ~6.5 mJ/cm$^2$ (blue circles). The blue line is a fit with the two fluid model (see text), while the black line indicates the Drude component of the two-fluid fit, accounting for the contribution of hot quasiparticles. The zero-frequency extrapolation of the latter, $\sigma_0^{QP}$, was used to estimate the equivalent quasiparticle temperature by comparison with the $\sigma_0^{QP}$ values obtained from the equilibrium conductivities measured at different temperatures [i,ii,iii]. The calibration scale obtained by this procedure is shown in (b).

**REFERENCES (Supplemental Material)**